\documentstyle[12pt]{article}

\def\beq{\begin{equation}}
\def\dd{{\cal D}}
\def\eeq{\end{equation}}

\def\hh{{\cal H}}
\def\hnu{H_{\nu}^{(1)}}
\def\hnd{H_{\nu}^{(2)}}

\def\mm{{\cal M}}
\def\nn{{\cal N}}
\def\lc{\left[}
\def\lp{\left(}
\def\rc{\right]}
\def\rp{\right)}
\def\sn{{\cal S}_N}
\def\vn{{\cal V}_N}
\def\vpk{\varphi_k}
\def\vpkinp{\varphi_{in,k}^{(+)}}

\def\vpkoutp{\varphi_{out,k}^{(+)}}
\def\vpkoutm{\varphi_{out,k}^{(-)}}
\def\vpkoutpm{\varphi_{out,k}^{(\pm)}}
\def\vv{{\cal V}}
\def\ws{\Sigma}
\def\zr{\zeta_R}
\def\zz{\zeta}

\begin{document}
\pagestyle{empty}
$\ $
\vskip 1.5 truecm

\centerline{\bf NUCLEATION RATES IN FLAT AND CURVED SPACE}
\vskip .75 truecm
\centerline{Jaume Garriga}
\vskip .4 truecm
\centerline{\em Tufts Institute of Cosmology,
              Department of Physics and Astronomy,}
\centerline{Tufts University, Medford, MA 02155}

\vskip 1. truecm

\begin{abstract}

Nucleation rates for tunneling processes in Minkowski and de Sitter space
are investigated, taking into account one loop prefactors.
In particular, we consider
the creation of membranes by an antisymmetric tensor field,
analogous to Schwinger pair production. This can be viewed as a model for
the decay of false (or true) vacuum at zero temperature, in the thin wall
limit. Also considered is the spontaneous nucleation of
strings, domain walls and monopoles during inflation.
The instantons for these processes are spherical world-sheets or world-lines
embedded in flat or de Sitter backgrounds. We find the contribution
of such instantons to the semiclassical partition function, including the
one loop corrections due to small fluctuations around the spherical
worldsheet. We suggest a prescription for obtaining, from the partition
function, the distribution of objects nucleated during inflation. This can
be seen as an extension of the usual formula, valid in flat space, according
to which the nucleation rate is twice the imaginary part of the free energy.
In addition, we use the method of Bogolubov transformations to compute
the rate of pair production by an electric field in 1+1 dimensional de Sitter
space, and compare the results to those obtained using the instanton method.
Both results agree where they are expected to, not only in the exponential
dependence but also in the prefactor, confirming the validity of instanton
techniques in de Sitter space. Throughout the paper, both the gravitational
field and the antisymmetric tensor field are assumed external.

\end{abstract}

\clearpage
\pagestyle{plain}
\section{Introduction}
\label{introduction}

A wide class of
non-perturbative phenomena in field theory can be understood in
terms of quantum tunneling. A well known example is the decay of false vacuum:
the materialization of bubbles of true vacuum in first order phase transitions
\cite{voal74,co85}. Lower dimensional versions of this process have
been used to model the decay of metastable topological defects,
such as domain walls and strings \cite{prvi92}.

A closely related phenomenon is the neutralization of the cosmological constant
through membrane creation.
In a spacetime of dimension $d=N+1$, an antisymmetric tensor
field $A$ of rank $N$ induces a cosmological constant. This is because the
corresponding field strength $F=dA$ has only one independent component,
which has to be constant in the absence of sources. Just as an electric field
decays through Schwinger pair creation, this cosmological constant decays
through membrane creation if $A$ is coupled to a membrane,
a process first described by Brown and Teitelboim
\cite{brte88}.

Such tunneling processes can happen in flat as well as in curved spacetime.
In addition, in curved spacetime, new effects can arise.
It has been shown that topological defects such as
circular loops of string,
spherical domain walls and monopole-antimonopole pairs can spontaneously
nucleate during inflation in the early universe
\cite{baal91}. These nucleations are
somewhat analogous to particle production by an external gravitational
field. Another consequence of space-time curvature is the possibility of
true vacuum decay \cite{lewe87}, through nucleation of {\em false} vacuum
bubbles.

Nucleation processes can be described using the
instanton methods \cite{co85}. The instantons are classical
solutions of the Euclidean equations of motion, with appropriate
boundary conditions. They are saddle points of the Euclidean path integral,
and as such they provide the basis for a semiclassical evaluation of the
partition function. The contribution of one instanton to the path integral
has the form
\beq
Ae^{-S_E},\label{form}
\eeq
where $S_E$ is the Euclidean action of the instanton, and the prefactor $A$
arises from Gaussian integration over small fluctuations around the
instanton. The main part of this paper will be devoted to the calculation of
the prefactors $A$ for the class of processes mentioned above.

The instanton methods can be applied in flat and in curved backgrounds.
One limitation of the formalism is, however, that the spacetime
under consideration has to have real Euclidean sections. Here, we shall
concentrate on de Sitter and Minkowski space. One
reason for studying de Sitter space is that it describes the geometry of
spacetime during inflation.
A de Sitter space of dimension $d$ can be defined as a hyperboloid
embedded in a Minkowski space of dimension $d+1$,
\beq
\eta_{AB}X^AX^B=H^{-2}.\label{hyperboloid}
\eeq
Here $X^A$ are the coordinates in the embedding Minkowski space
($A=0,...,d$), $H$ is the expansion rate during inflation and $\eta_{AB}$ is
the Minkowski metric. We use the metric convention $(-,+,...,+)$.
The Euclidean section of de Sitter space is obtained by analytically
continuing the temporal coordinate $X^0$ to imaginary values
$$
X^0=-iX^0_E
$$
(with $X_E^0$ a real number). With this rotation the hyperboloid
(\ref{hyperboloid}) becomes a $d$-sphere of radius $H^{-1}$ embedded in
flat Euclidean space (see Fig. 1).

In flat space, at zero or at finite temperature, the nucleation rates can
be related to the imaginary part of the free energy \cite{af81,co85}, and
they are essentially given by an expression of the form (\ref{form}). In
curved space, it is believed that the nucleation rates also have the
exponential dependence (\ref{form}), although the theory has not been
developed to the same level of rigor than in flat space.
When the size of the instantons is very small compared with $H^{-1}$, one
expects that the usual flat space formulas should apply. However, for the
spontaneous nucleation of topological defects and for the nucleation of
{\em false} vacuum bubbles, the size of the instanton is comparable to the
horizon scale $H^{-1}$. In such cases it is not clear how one should
compute the nucleation rate, and one may even question whether such
tunnelings can occur.

In this paper we take the heuristic point of view that these nucleations can
indeed occur. We suggest a prescription for computing,
from the semiclassical partition function, the
equilibrium distribution of created membranes, bubbles and topological
defects during inflation. This can be seen as a generalization of the formulas
that one uses in flat space.

In 1+1 dimensions the process of membrane (or bubble) creation reduces to
that of particle creation in an external field. In that case,
the predictions of the instanton method can be compared with the results
that one obtains by using the better understood method of Bogolubov
transformations. As we shall see, the results of both methods agree (in the
limit where they are valid), even in the case when the size of the
instanton is comparable to the horizon scale. Then, at least in 1+1
dimensions, the instanton prescription seems to be valid, and there is no
reason to believe that it will not be valid in higher dimensions.

Finally, we should mention that throughout the paper, both the
gravitational and the antisymmetric tensor fields are asumed
to be external. As we shall see, the backreaction of the membrane on
the antisymmetric field can be neglected when the charge $e$ of the membrane
is very small compared to the field strength.
In the context of false
vacuum decay, this means that the difference in energy density between
the vacua in the two sides of the membrane
will be much smaller than the overall
cosmological constant. Also, in order that gravitational field of the
membrane be negligible, the mass scale of the membrane should be sufficiently
small (see e.g \cite{baal91} for a comparison of instantons with and
without self gravity). Gravitational backreaction effects are
interesting in their own right, leading to qualitatively different
behaviour at large mass scales, but introducing new complications and
problems \cite{LRT85,tasa92}. These will be left as subject for future
research.

The plan of the paper is the following. In section 2 we study the
instantons for membrane production. These are the same as the ones
studied in Ref. \cite{brte88}, in the limit of negligible self gravity and
small charge. These instantons are essentially spherical
Euclidean worldsheets of dimension $N$ and radius $R_0$
(representing the membrane) embedded in
Euclidean de Sitter space, which is itself a sphere of radius $H^{-1}$ and
dimensionality $d=N+1$. The radius $R_0$ is determined by the
strength of the antisymmetric
field, the charge of the membrane, its surface tension and the expansion
rate $H$. Special attention is paid to the analytic continuation of the
instantons to Lorentzian signature, and the effect of de Sitter
transformations on the resulting solutions. These
Lorentzian solutions describe the motion of the membranes after nucleation.

As mentioned above, to calculate the prefactor $A$ we need to integrate
over small fluctuations around the instanton. In Section 3 the theory of
such fluctuations is reviewed, with an emphasis in the so-called zero
modes. The zero modes are perturbations which do not change the shape of
the instanton, but correspond to infinitessimal translations of the
solution as a whole. We make use of the covariant formalism developed in
Refs. \cite{gavi91,gavi92,jemal1}, according to which the worldsheet
fluctuations are represented by a scalar field $\phi$ `living' in the
unperturbed worldsheet. This scalar field has the meaning of a
normal displacement of the worldsheet.

In Section 4, we briefly review
the instantons for the spontaneous
nucleation of topological defects during inflation. These can be seen as a
limiting case of the ones for membrane creation, when the
external antisymmetric field is switched off. However, in this case the
co-dimension of the defect's worldsheet can be larger than one and
additional perturbations have to be considered. For later use in the paper,
we also discuss the instantons for a massive particle at finite temperature
and for pair production in 3+1 dimensions. We study fluctuations around
these instantons and give the normalization of the zero modes.

Section 5 is devoted to the semiclassical evaluation of the partition
function. It is shown that the evaluation of the prefactor is formally
equivalent to the evaluation of the effective action for a free scalar
field in curved space-time (this curved space-time is the worldsheet of the
instanton; a sphere, in our case).
To illustrate the method, the instanton formalism is used to
compute the partition function for a gas of massive particles at finite
temperature, in the semiclassical limit $\mm>>T$, where $\mm$ is the
particle's mass and $T$ is the temperature.

In section 6 we compute nucleation rates in flat space, recovering known
results for pair creation in 1+1 and 3+1 dimensions, and for bubble
formation in 2+1 and 3+1 dimensions. The question of renormalization is
briefly discussed in analogy with the renormalization of the effective
action for a scalar field in curved space.

In section 7 we discuss the nucleation rates in curved spacetime. We find
the size distribution of created membranes, bubbles and defects
during inflation. We also give the momentum distribution for the case of
pair creation.

In section 8 we study the quantization of a charged scalar field
interacting with an external electric field in 1+1 dimensional de Sitter
space. The spectrum of particles created by the electric and gravitational
field is computed using the method of Bogolubov transformations.

Some conclusions are summarized in Section 9. The computation of functional
determinants on the N-sphere, necessary for the evaluation of the
instanton prefactors, is done in the Appendix.

\section{Production of membranes by an antisymmetric tensor field.}

In this section we describe the instantons for the creation of membranes
by an antisymmetric tensor field and their analytic continuation to
Lorentzian signature. The instantons discussed in subsection
\ref{instantonsub} also represent the formation of true (or false) vacuum
bubbles in the thin wall limit.

\subsection{Particle coupled to an electric field in 1+1 dimensions}

For the main part of the paper, the antisymmetric
tensor field will be assumed external. However, it is instructive
to start our discussion with the 1+1 dimensional case and treating the
electric field as dynamical.
The action for a spinless particle of mass $\mm$ an charge $e$
interacting with a Maxwell field $A_{\mu}$ in 1+1 dimensions is given by
\cite{brte88}
\beq
S=-\mm\int_{\ws}ds+e\int_{\ws}A_{\mu}dx^{\mu}-{1\over
4}\int d^2x\sqrt{-g}F_{\mu\nu}F^{\mu\nu}+\int
d^2x\partial_{\mu}(\sqrt{-g}F^{\mu\nu}A_{\nu}).
\label{brte}
\eeq
Here $x^{\mu}(s)$ is the particle's trajectory and
$ds$ is the proper time interval. The particle's
world-line is indicated by $\ws$ and $g$ denotes the determinant of the
spacetime metric $g_{\mu\nu}$. The last term is a boundary term, included
to yield a well defined equation for $A_{\mu}$. (This is
necessary since we shall consider situations in which the field strength
$F_{\mu\nu}=\partial_{\mu}A_{\nu}-\partial_{\nu}A_{\mu}$ does not vanish at
infinity \cite{brte88}.)

Using the notation of forms, $A=A_{\mu}dx^{\mu}$, the field strength is
$F=dA$. In 1+1 dimensions $F$ has only one independent component, the
electric field $E$,
\beq
F=-E\tilde \epsilon\label{one}
\eeq
where
\beq
\tilde\epsilon=\sqrt{|g|}\epsilon_{\mu\nu}dx^{\mu}\wedge dx^{\nu}.
\label{etilde}
\eeq
 Eq.(\ref{one}) is just saying that in 1+1 dimensions any
2-form is proportional to the ``volume'' form $\tilde\epsilon$.
(Here $\epsilon_{\mu\nu}$ is the antisymmetric symbol, with
$\epsilon_{01}=1$.)

Variation of (\ref{brte}) with respect to $A_{\mu}$ yields the Maxwell
equation
\beq
\partial_{\mu}(\sqrt{-g}F^{\mu\nu})=-e\int ds\delta^2(x-x(s))
{dx^{\nu}(s)\over ds}.
\label{max}
\eeq
This equaton implies that in the absence of sources the electric field
is constant
$$
E=const.
$$
whereas the effect of a charged particle is to produce a discontinuity
$$
\Delta E=e
$$
in the electric field as we cross from one side of the worldline to the
other. Therefore, the only equation for $A_{\mu}$ is Gauss' law, which
is an equation of constraint \cite{co76}. In this sense
$A_{\mu}$ does not actually have any field degrees of freedom in 1+1
dimensions.

As is well known, a constant electric field will decay by producing pairs
of charged particles, which will then accelerate away from each other,
screening the electric field away. The rate of pair production can be
computed using the instanton methods. For this, it is necessary to solve
the
Euclidean equations of motion. The Euclidean action $S_E$ can be found by
complexifying the temporal coordinate $x^0\to-ix^0_E$, $ds\to-ids_E$ and
leaving the field strength $F^{\mu\nu}\to F^{\mu\nu}_E$ unchanged
\cite{brte88}. From $F^{01}=\partial^{0}A^{1}-\partial^1A^0$ one has to
complexify the vector potential $A^0\to A^0_E$, $A^1\to iA^1_E$. With such
rotations the Euclidean action is found to be ($iS\to -S_E$),
\beq
S_E=+\mm\int_{\ws}ds+e\int_{\ws}A_{\mu}dx^{\mu}-{1\over4}\int
d^2\sqrt{g}F_{\mu\nu}F^{\mu\nu}+
\int dx^2 \partial_{\mu}(\sqrt{-g}F^{\mu\nu}A_{\nu}).\label{euc}
\eeq
Here we have dropped the Euclidean subscripts. Expanding the derivative in
the last term and using (\ref{max}) one has
$$
S_E=\mm\int_{\ws}ds+{1\over 4}\int d^2x\sqrt{g}F_{\mu\nu}F^{\mu\nu},
$$
which is positive definite.

Unlike the Lorentzian case, in Euclidean space we can consider closed
worldlines, and actually these will be the ones relevant to pair production. A
closed wold-line divides spacetime into two regions. Following
\cite{brte88} we denote them as the ``inside'' and the ``outside''
of the worldline (see Fig. 1). Note that in flat
space there is a natural way to assign these labels, but in de Sitter space
(that is on the 2-sphere) these denominations are just conventional. We
denote by ``outside'' the region in which, upon analytic continuation,
the electric field does
not change when the pair is created,
maintaining its initial value $E_0$. By Gauss' law, the field in
the inside region is given by $E_i=E_0\pm e$. The double sign reflects the
fact that the worldline has to be assigned a direction in which the charge
flows, and in principle both directions are possible. Without loss of
generality, we take the minus sign,
\beq
E_i=E_0-e,\label{ei}
\eeq
since a change in the direction of the worldline amounts
to a change in the sign of $e$.

Using $F_{\mu\nu}F^{\mu\nu}=2E^2$, we have
\beq
S_E = \mm\int_{\ws}ds-(eE_0-{e^2\over 2})\int_{\vv}\tilde
\epsilon.\label{same}
\eeq
Here $\vv$ is the volume of the ``inside'' region. In (\ref{same}) we have
dropped the constant term
$$
{1\over 4}\int d^2x\sqrt{g} 2E_0^2.
$$
This term is just the Euclidean action for the background configuration,
without the instanton. Of course this constant is physically meaningless
because a constant can always be added to the action. It is customary to
choose the constant so that the action for the background
configuration vanishes \cite{co85}.
To summarize, the action (\ref{same}) is proportional
to the length of the worldline minus a term proportional to the area
enclosed by the worldline.

\subsection{Antisymmetric tensor field coupled to a membrane}

Let us now consider the natural generalization of (\ref{brte}), which
describes an antisymmetric tensor field of rank $N$
$$
A=A_{[\mu...\rho]}dx^{\mu}\wedge...\wedge dx^{\rho}
$$
interacting with the $N$
dimensional worldsheet of a charged membrane
in a spacetime of dimension $d=N+1$ \cite{brte88}. To keep the
discussion simple, we shall consider the situation in which the field $A$
is external, so that backreaction of the membrane on the field is ignored.
This will facilitate also the comparison with the method
of Bogolubov coefficients in 1+1 space-time dimensions, since with that method
the electric field has to be treated as external anyway.

The action is now
given  by
\beq
S=-\mm\int_{\ws}\sqrt{-\gamma}d^N\xi+e\int_{\ws}A.\label{actionext}
\eeq
The first term is the Nambu action, proportional to the area of the worldsheet
$\ws$, where $\gamma$ is the determinant of the worldsheet
metric $\gamma_{ab}$, and
$\xi^a$ ($a=0,...,N-1$) is a set of
coordinates on $\ws$ .
In (\ref{actionext}) $\mm$ is just a constant. For $N$=1 this
constant is the particle mass, for $N$=2 it is the tension of a string,
and for $N=3$ it is the surface tension of a membrane.
The second term in (\ref{actionext}) is the
generalization of the electromagnetic coupling $e\int A_{\mu}dx^{\mu}$
that we used in (\ref{brte}).
Eq. (\ref{actionext}) can be Euclideanized using prescriptions
analogous to the ones
that led from (\ref{brte}) to (\ref{euc}), yielding
\beq
S_E=\mm\int_{\ws}\sqrt{\gamma}d^N\xi+e\int_{\ws}A.\label{eucex}
\eeq
As with the 1+1 dimensional case, the field strength associated with $A$,
$F=dA$, has only one independent component, and can be written in the form
(\ref{one}), where now the antisymmetric symbol $\epsilon$ has $N+1$
indices.

For closed worldsheets we can use Stokes' theorem and (\ref{one}) in
(\ref{eucex}) to find
\beq
S_E=\mm\int_{\ws}\sqrt{\gamma}d^N\xi-eE_0\int_{\vv}\tilde \epsilon,
\label{mohi}
\eeq
where we have used a constant external electric field $E_0$. Without loss
of generality we take
$$
E_0>0.
$$
The integral in the last term is just the volume of the space-time region
``inside'' the closed worldsheet. Eq. (\ref{mohi}) has the same form as
(\ref{same}) if the term proportional to $e^2$ is neglected. Therefore one
can neglect backreaction when $|e|<<E_0$.

Notice that (\ref{mohi}) is proportional to the area of the worldsheet
minus a term proportional to the volume enclosed by the worldsheet. This
has exactly the same form as the Euclidean action for the process of false
vacuum decay through bubble nucleation \cite{co85}, in the limit in which
the thickness of the wall separating the true from the false vacua is much
smaller than the radius of the bubble. Both processes are similar in many
respects \cite{brte88}, the main difference being that membrane production
by an antisymmetric tensor field can occur repeatedly at any given point in
space, whereas vacuum decay occurs only once.

\subsection{The instantons}
\label{instantonsub}

The equation of motion following from (\ref{mohi}) has been given for
instance in Refs \cite{gavi91,jemal1}
\beq
K^a_a=\gamma^{ab}K_{ab}=-{eE_0\over\mm}.\label{constantext}
\eeq
where $K_{ab}$ is the extrinsic curvature of the Euclidean worldsheet
\beq
K_{ab}\equiv -e_{b\mu}\nabla_a n^{\mu}. \label{extri}
\eeq
Here $n^{\mu}$ is the normal to the worldsheet, and
$e_b^{\mu}=\partial_b x^{\mu}(\xi^c)$ are the tangent vectors
(our sign convention is that $n^{\mu}$ points towards the outside region.)

In 1+1 dimensional flat spacetime, the only solution of (\ref{constantext})
is a circular worldline of radius $R_0=\mm/eE_0$,
\beq
(x^0_E)^2+(x^1)^2=R_0^2.\label{circle}
\eeq
Since we have chosen $E_0>0$, in order for $R_0$ to be positive we need
$$e>0.$$
{}From (\ref{ei}), this means that the electric field inside the circle will
be smaller than outside.

The evolution of the pair after nucleation is given by the analytic
continuation of (\ref{circle}) to Minkowski space
\beq
-(x^0)^2+(x^1)^2=R_0^2.\label{hy11}
\eeq
This hyperbola has two branches. The branch on the right represents a
particle of charge $e$ moving forward in time. The one on the left
represents a particle of charge $e$ moving backward in time. This is
interpreted in the usual way as an antiparticle of charge $-e$ moving
forward in time.
The particle and antiparticle pair nucleate at time $x^0=0$, separated by a
distance $2R_0$ and with zero velocity. After that, due to the constant
force exerted by the field, they start moving away from each other with
constant proper acceleration $R_0^{-1}$.

In higher dimensions there are also $N-$spherical worldsheets which are
extrema of the action. These represent the nucleation of spherical
membranes \cite{brte88}.
Let us consider directly the instantons in de Sitter space
of radius $H^{-1}$.
The flat space instantons can be obtained as the limiting case $H\to 0$.

With a spherical ansatz for
the worldsheet the action (\ref{mohi}) takes the form
\beq
S_E=\mm\sn(R_0)-eE_0\vn(\theta_0).\label{actionsph}
\eeq
Here
\beq
\sn(R_0)={2\pi^{N+1\over 2}\over\Gamma\left({N+1\over 2}\right)}R_0^N
\label{sn}
\eeq
is the surface of a worldsheet of radius $R_0$, $\theta_0$ is the
polar angle on the $d$-sphere of radius $H^{-1}$ (see Fig. 1) such that
$$
R_0=H^{-1}\sin \theta_0,
$$
and $\vn(\theta_0)$ is the volume of the $d$-sphere that is enclosed by
the worldsheet of radius $R_0$:
\beq
\vn(\theta_0)={2\pi^{N+1\over 2}\over\Gamma\left({N+1\over 2}\right)}H^{-(N+1)}
\int_0^{\theta_0}\sin^N\theta d\theta.
\label{vn}
\eeq
Extremizing (\ref{actionsph}) with respect to $\theta_0$
$$
{dS_E\over d\theta_0}=0
$$
we find
\beq
\tan\theta_0=NH{\mm\over eE_0},\label{tan}
\eeq
which means that the radius of the Euclidean worldsheet is
\beq
R_0={N\mm\over(N^2H^2\mm^2+e^2E_0^2)^{1/2}}.\label{r0}
\eeq
Substituting (\ref{r0}) back into (\ref{actionsph}) we find the Euclidean
action for the instantons
\beq
S_E=2\pi H^{-2}\left[(\mm^2H^2+e^2E_0^2)^{1/2}-eE_0\right]\quad\quad(N=1)
\label{se1}
\eeq
\beq
S_E=4\pi H^{-3}\left[\mm H-{eE_0\over 2}\arctan{2H\mm\over eE_0}\right]
\quad\quad(N=2) \label{se2}
\eeq
\beq
S_E={2\over 3}\pi^2H^{-4}\left[{9H^2\mm^2+2e^2E_0^2\over
(9H^2\mm^2+e^2E_0^2)^{1/2}}-2eE_0\right].\quad(N=3)\label{se3}
\eeq
An interesting feature of Equations (\ref{se1}-\ref{se3}) is that one finds
instantons of finite action both for $e>0$ and $e<0$.

For $e>0$ the electric field in the inside region decreases with respect
to the initial value [see(\ref{ei})]. From (\ref{tan}) this corresponds to
$\theta_0<{\pi\over 2}$. For $e<0$ the electric field in the inside region
actually {\em increases} with respect to the initial value. This corresponds to
$\theta_0>{\pi\over 2}$. In this case the ``inside'' region is
actually larger than the ``outside'' one (see Fig. 1). Strictly speaking, if
the electric field is treated as external, we should not say that the field
increases or decreases in the inside region. However, both cases should still
be distinguished. For instance, in 1+1 dimensions, the instanton with $e>0$
corresponds to the creation of a pair with the ``screening'' orientation. That
is, after nucleation, the + charge is to the right of the inside region and
the $-$ charge is to the left (recall our convention $E_0>0$). On the other
hand, for $e<0$ the pair has the ``anti-screening'' orientation, with the
+ charge to the left and the $-$ charge to the right. Similarly membranes
can nucleate with two different orientations depending on the sign
we take for $e$.

It might appear that the particles in pairs with the anti-screening
orientation would move towards each other after nucleation, and eventually
anihilate each other. However, as we shall see, the distance between both
particles actually grows with time due to the inflationary expansion.
We should emphasize that the anti-screening instantons are just as
physical as the screening ones. In order to find agreement
with the results obtained using Bogolubov transformations, both instantons
will have to be included. In the context of vacuum decay, the case $e>0$
corresponds to the ordinary transition from false to true vacuum, whereas
the case $e<0$ corresponds to the decay of the true vacuum through nucleation
of false vacuum bubbles \cite{lewe87}.

In the limit when the electric field is switched off, $E_0\to 0$,
the instantons become spheres of maximal radius $R_0\to H^{-1}$, and the
action (\ref{se1}-\ref{se3}) reduces to
\beq
S_E=\mm\sn(H^{-1}).\label{senoelec}
\eeq
These are the instantons for the spontaneous nucleation of defects
during inflation \cite{baal91}, which we shall consider in more detail
later on.

In general, for finite $E_0$, the action for $eE_0>0$ is always
smaller than that for $eE_0<0$.
This means that it is more probable to nucleate a screening
membrane than an anti-screening one, in agreement with naive expectations.

In the flat space limit $H\to 0$, the anti-screening process has infinite
action, and therefore it is not possible.
Only the action for the screening instanton $e>0$ remains finite. For $e>0$
and $H\to 0$ we have
\beq
R_0={N\mm\over eE_0}
\label{roflat}
\eeq
and
\beq
S_E={\mm\over N+1}\sn(R_0).\label{seflat}
\eeq
This expression reproduces the thin wall instanton action for
vacuum decay in flat space \cite{co85}, where $\mm$ is the tension of the
wall and $eE_0$ is the difference in energy density between the false and true
vacua.

\subsection{Analytic continuation}

The evolution of the membranes after nucleation is
given by the analytic continuation of the instantons back to Lorentzian
signature. We have seen that, for the $d=1+1$ case in flat space the
instanton is a circle, and the analytic continuation is a hyperbola
representing the world-line of the particle and antiparticle accelerating
away from each other. Note that the hyperbola (\ref{hy11}) is centered at
the origin $x^1=0$, but of course pairs can nucleate at other locations
too. If we act on the instanton (\ref{circle}) with a space-time
translation, the resulting trajectory
$$
(x^0_E-a_E)^2+(x^1-b)^2=R_0^2
$$
is also an instanton. Thus (\ref{circle}) is just one solution out of a two
parameter family. By analytically continuing $x^0_E\to i x^0$
and $a_E\to i a$, we obtain a two parameter family of Lorentzian solutions
\beq
-(x^0-a)^2+(x^1-b)^2=R_0^2,\label{lloro}
\eeq
which represent pairs nucleating at any space-time point $x^{\mu}=(a,b)$.

Similar steps have to be taken to analytically continue the
instantons describing the creation of membranes in de Sitter.
As mentioned before these are $N-$spherical worldsheets of radius $R_0$
embedded in
d-sphere of radius $H^{-1}$. The instantons can be represented as the
intersection of the $d-$sphere with a hyperplane at a distance
$$
\omega_0\equiv H^{-1}\cos\theta_0
$$
from the origin, where $\theta_0$ is given by (\ref{tan}), see Fig. 1.
In the
representation (\ref{hyperboloid}), the instanton is given by
$$
(X^0_E)^2+\sum^d_{J=1}(X^J)^2=H^{-2}
$$
\beq
X^d=\omega_0,\label{instmem}
\eeq
where $X^A$ are the coordinates in a `fictitious' embedding Euclidean
space.

Following \cite{baal91}, to analytically continue this solution we choose
the flat Friedman-Robertson-Walker (FRW) coordinates in de
Sitter space, $(t,\vec x)$. In these, the metric takes the form
\beq
ds^2=-dt^2+e^{2Ht}d\vec x^2.\label{metric}
\eeq
These coordinates are related to $X^A$ through the relations
$$
X^0=H^{-1}\sinh Ht+ {1\over 2}H\vec x^2e^{Ht},
$$
\beq
X^d=H^{-1}\cosh{Ht}-{1\over 2}H\vec x^2e^{Ht},\label{transform}
\eeq
$$
\vec X=\vec x e^{Ht},
$$
where the vector $\vec X$ has components $X^J$, $(J=1,...,d-1)$.
The coordinates $(t,\vec x)$ cover only half of the hyperboloid
(\ref{hyperboloid}).

Taking $X^d=\omega_0$ in (\ref{transform}), the worldsheet of the membrane
after nucleation is given by
\beq
\vec x^2=H^{-2}(1+e^{-2Ht})-2H^{-1}\omega_0e^{-Ht}.\label{wsa}
\eeq
This solution represents a spherical membrane which is expanding in time,
with physical radius given by
\beq
R^2=H^{-2}(e^{2Ht}+1)-2H^{-1}\omega_0e^{Ht}.\label{rphyswsa}
\eeq
Note that sign($\omega_0$)=sign($e$), but $R$ never vanishes for either
sign of $e$. In both cases, the radius grows like the scale factor at
late times.
For $N=1$ the spherical ``membrane'' reduces to a pair of
points, whose worldline is given by $x=\pm (\vec x^2)^{1/2}$, with $\vec
x^2$ given by (\ref{wsa}).
In this case $R$ is one half of the physical
distance between the particle and antiparticle in the pair.

As with the flat space case discussed above, the solution (\ref{wsa})
belongs to a family of solutions which can be obtained from a $d$-parameter
family of instantons. This family is obtained by applying $O(d+1)$
rotations to the instanton (\ref{instmem}). The group $O(d+1)$ has
$d(d+1)/2$ generators. Of these, $d(d-1)/2$ leave the instanton invariant.
They correspond to rotations in the space $(X^0_E,\vec X)$. The remaining
$d$ generators correspond to rotations in the $(X^d,X^0_E)$ plane or in any
of the $(X^d,X^I)$ planes $(i=1,...,d-1)$. These generators which do {\em
not}
leave the instanton invariant are the so called zero modes. Their effect is
to rotate the hyperplane $X^d=\omega_0$ in (\ref{instmem}), effectively
translating
the center of the worldsheet to a new location on the $d$-sphere.

Upon analytic continuation
the parameters
corresponding to rotations in the $(X^0_E,X^J)$ plane $(J\neq 0)$
have to be complexified along with $X^0_E$ in order for the resulting
solutions to be real. Recall that even in flat space, the parameter $a_E$
had to be complexified to obtain (\ref{lloro}). In the present case,
rotations turn into boosts in the $(X^0,X^J)$ plane when the angle
$\alpha_E$ of rotation is complexified,
\beq
\alpha_E=i\alpha.\label{iguana}
\eeq
In this way,
the group of rotations $O(d+1)$ becomes the
group of de Sitter transformations $O(d,1)$, which can
also be thought of as the
group of Lorentz transformations in the Minkowski space in which the
hyperboloid (\ref{hyperboloid}) is embedded.

Let us consider the case $d=1+1$ in some detail.
The general worldline after
nucleation is obtained by taking $X^0_E\to iX^0$ in
(\ref{instmem}) and then applying a
boost in the $(X^0,X^2)$ plane followed by a rotation in the $(X^1,X^2)$
plane
\beq
X'^0=X^0\cosh\alpha+X^2\sinh\alpha \label{general}
\eeq
$$
X'^1=-(X^0\sinh\alpha+X^2\cosh\alpha)\sin\beta+X^1\cos\beta
$$
$$
X'^2=(X^0\sinh\alpha+X^2\cosh\alpha)\cos\beta+X^1\sin\beta.
$$
Here $\alpha$ and $\beta$ are arbitrary parameters.

Elliminating $X^1$
and $X^0$ from the previous equations one has
$$
X'^2\cosh\alpha\cos\beta-X'^1\cosh\alpha\sin\beta-X'^0\sinh\alpha=X^2.
$$
Taking $X^2=\omega_0$, dropping primes and using the transformations
(\ref{transform}) we find, after some algebra, that the general world-line
after nucleation is given by
\beq
R^2\equiv
e^{2Ht}(x-x_0)^2=H^{-2}(e^{2H(t-t_0)}+1)-2H^{-1}\omega_0e^{H(t-t_0)},
\label{genwl}
\eeq
where
\beq
x_0=H^{-1}{\cosh\alpha\sin\beta\over\cosh\alpha\cos\beta+\sinh\alpha}
\label{x0w}
\eeq
\beq
t_0=H^{-1}\ln(\cosh\alpha\cos\beta+\sinh\alpha).\label{t0w}
\eeq
Eq. (\ref{genwl}) represents a pair centered at the point $x_0$. The
parameter $t_0$ shall be refered to as the time of nucleation.
For $R_0\sim H^{-1}$ this is somewhat conventional, since there is no precise
instant of time at which the pair nucleates \cite{baal91}. The problem is that
the concept of simultaneity becomes blurry at distances comparable to the
horizon. Strictly speaking, all we can
say is that solutions with different values of $t_0$ are time translations
of one another. From a geometric point of view, however, it is clear that
$(x_0,t_0)$ represents the center of symmetry of the worldsheet, and in
this sense it is natural to think of it as the nucleation event.

Note also that there is no absolute value sign in the logarithm
in (\ref{t0w}). For $\omega_0\neq0$, the
solutions with $\cosh\alpha\cos\beta+\sin\alpha>0$ are qualitatively
different from the ones with $\cosh\alpha\cos\beta+\sinh\alpha<0$.
Actually, we shall see that the latter ones are unphysical. They correspond
to pairs whose ``inside'' region is centered at spatial infinity, so upon
nucleation the electric field would change over an infinite (and
disconnected) region of space. To find agreement with the Bogolubov method
these solutions have to be discarded.

The same arguments can be repeated for $N>1$.
The general Lorentzian solution is a spherical membrane of physical radius
(\ref{rphyswsa}) centered at any spacetime point \cite{baal91}.

\section{Perturbations and zero modes}
\label{perturbations}

To compute the contribution of the instantons to the partition
function we need to study small fluctuations of the instanton
worldsheet. For this it is very useful to adopt
the covariant formalism developed in Refs.
\cite{gavi91,gavi92,jemal1,jemal2}, according to which the
worldsheet perturbations are represented by a scalar field
``living'' on $\ws$, which has the meaning of normal displacement.

Denoting by $x^{\mu}$ the coordinates in de Sitter space,
the instanton configuration
will be denoted by $x^{\mu}(\xi^a)$. Consider now a slight deformation of
the worldsheet
\beq
\tilde x^{\mu}(\xi^a)=x^{\mu}(\xi^a)+\delta x^{\mu}(\xi^a).\label{pws}
\eeq
Since only
deformations orthogonal to the worldsheet are physically meaningful,
we can set
\beq
\delta x^{\mu}(\xi^a)= \mm^{-1/2}\phi(\xi^a)n^{\mu}.\label{phinmu}
\eeq
Here $n^{\mu}$ is the normal to the worldsheet and $\phi(\xi^a)$ has
the meaning of a normal displacement.
The factor $\mm^{-1/2}$ is inserted so that $\phi$ has the
correct dimensions for a scalar field in $N$ dimensions,
$[\phi]=(mass)^{(N/2)-1}$.

Actually, the equation of motion for $\phi$ can be derived
from kinematical considerations.
Since $\phi$ is a scalar on the worldsheet, it has
to satisfy a covariant equation. The only tensors available in
$\ws$ are the metric $\gamma_{ab}$ and the extrinsic curvature $K_{ab}$,
but because of the symmetries of our problem they are proportional to each
other
$K_{ab}\propto\gamma_{ab}$ (see e.g. \cite{baal91}). The only covariant
second order differential equation that we can write down with such
ingredients is
\beq
-\Delta \phi+M^2\phi=0,
\label{lap}
\eeq
where $\Delta$ is the Laplacian on the spherical worldsheet.
By symmetry, $M$ has to be constant.

To determine the value of $M$ we can use ``known'' solutions of
(\ref{lap}): the zero modes. These are field modes for $\phi$ which do not
correspond to true perturbations of $\ws$, but to the infinitessimal version
of the
rotations considered in the previous section. These solutions can be
found from geometric considerations. Let the instanton be given by
[see (\ref{instmem})]
\beq
(X^0_E)^2+\sum_{J=1}^d(X^J)^2=H^{-2}\label{021}
\eeq
$$
X^d=\omega_0=(H^{-2}-R_0^2)^{1/2},
$$
where, as usual, $X^A$ are cartesian coordinates in the flat space in
which the Euclidean de Sitter space is embedded.

The vector $n^{\mu}$, orthogonal to $\ws$, can be thought of as a vector
in this embedding flat space, with components $n^A$. As such, it is tangent
to the $d-$sphere of radius $H^{-1}$ centered at the origin, and orthogonal
to the  $N-$sphere of radius $R_0$ centered at $X^d=\omega_0$, $X^I=0$ $(I\neq
d$). It is easy to see that the normal vector is given by
$$
n^A=HR^{-1}_0(\vec X\omega_0,-R_0^2).
$$
Here $\vec X$ has components $X^I$, $(I\neq d)$. A small
rotation of angle $\alpha$ in the $(X^J,X^d)$ plane induces the change
$$
\delta X^J=\alpha X^d,
$$
$$
\delta X^d=-\alpha X^J.
$$
This transforms $\ws$ into a new worldsheet which is also a solution of the
equations of motion. Therefore, taking $X^d=\omega_0$ in the equations above,
the field
\beq
\mm^{-1/2}\phi(\xi^a)\equiv n^A\delta X_A(\xi^a)=\alpha(HR_0)^{-1}X^J(\xi^a)
\label{zirs}
\eeq
has to be a solution of (\ref{lap}), for any $J=0,...,d-1$.
As is well known,
the cartesian components $X^J(\xi^a)$ of the points on the $N-$sphere
are linear
combinations of the spherical harmonics with $L=1$. The spherical harmonics
are eigenfunctions of the Laplacian with eigenvalue
\beq
\lambda_L=-L(L+N-1)R_0^{-2},\quad (L=0,...,\infty) \label{lambdal}
\eeq
so, taking $L=1$, we have $\Delta\phi=-NR_0^{-2}\phi.$ Comparing with
(\ref{lap}) we have the mass that we were looking for
\beq
M^2=-NR_0^{-2}. \label{m2}
\eeq
For $eE_0\to 0$ this reduces to $M^2=-NH^2$, a result which was found
already in \cite{baal91}.

Of course, equations (\ref{lap}) and (\ref{m2}) can also be derived
from a perturbative expansion of the action.
Introducing (\ref{pws}) in the action and expanding
to second order in $\phi$,  the
action for the perturbed worldsheet $\tilde\Sigma$ can be written as
\cite{gavi91,gavi92,jemal1},
$$
S_E[\tilde\Sigma]=\bar S_E[\ws]+S^{(2)}_E[\phi],
$$
where $\bar S_E[\ws]$ is the action for the unperturbed instanton
(\ref{se1}-\ref{se3}), and
\beq
S_E^{(2)}[\phi]={1\over 2}\int
d^N\xi\sqrt{\gamma}(-\phi\Delta\phi+M^2\phi^2),
\label{sphi}
\eeq
whith $\Delta$ the Laplacian on $\ws$ and
\beq
M^2={\cal
R}^{(N)}-R^{(d)}_{\mu\nu}(g^{\mu\nu}-n^{\mu}n^{\nu})-\left({eE_0\over\mm}
\right)^2.\label{m2comp}
\eeq
Here ${\cal R}^{(N)}$ is the Ricci scalar on $\ws$, and $R^{(d)}_{\mu\nu}$ and
$g_{\mu\nu}$ are the Ricci tensor and the metric in the embedding de Sitter
space [actually Eqs. (\ref{sphi}) and (\ref{m2comp}) are valid for
pertubations to any worldsheet solution,
embedded in an arbitrary curved spacetime of dimension $d=N+1$].
In de Sitter space (i.e. on the $d$-sphere)
$R^{(d)}_{\mu\nu}=H^2(d-1)g_{\mu\nu},$
whereas on the $N$-spherical worldsheet of radius $R_0$,
${\cal R}^{(N)}=\gamma^{ab}\ R^{(N)}_{ab}=N(N-1)R_0^{-2}.$
Substituting in (\ref{m2comp}) and using (\ref{r0}) for $R_0$, the
effective mass $M^2$ simplifies to
$$
M^2=-NR_0^{-2},
$$
in agreement with (\ref{m2}).

For later convenience we expand an arbitrary perturbation $\phi$ in terms
of the (real) spherical harmonics on the $N$-sphere, $\phi_{LJ}$
\beq
\phi(\xi^a)=\sum_{LJ}C_{LJ}\phi_{LJ}(\xi^a).\label{harmonex}
\eeq
These satisfy $\Delta \phi_{LJ}=\lambda_L\phi_{LJ}$, with $\lambda_L$ given
by (\ref{lambdal}) and
\beq
\int_{\ws}\phi_{LJ}^2\sqrt{\gamma}d^N\xi=1.
\label{norma}
\eeq
The index $J$ labels the degeneracy for given $L$
$$
J=0,...,g_L-1,
$$
where
\beq
g_L={(2L+N-1)(N+L-2)!\over L!(N-1)!}.
\label{gl}
\eeq
For $L=1$ we have, with the normalization (\ref{norma}),
\beq
\phi_{1J}=\left({N+1\over R_0^2\sn(R_0)}\right)^{1/2}X^J(\xi),
\label{sinyo}
\eeq
where $\sn$ is given by (\ref{sn}).

Comparing (\ref{zirs}) with (\ref{harmonex}) and using (\ref{sinyo})
we find that for an infinitessimal rotation
of angle $d\alpha_J$ in the $(X^J,X^n)$
plane
\beq
dC_{1J}=H^{-1}d\alpha_J\left({\mm\sn(R_0)\over N+1}\right)^{1/2}.
\label{normzir}
\eeq
This equation is often refered to as the normalization of the zero modes,
and it will be important in order to interpret certain divergences in the
semiclassical evaluation of the partition function. In the flat space
limit, the square root in the right hand side of (\ref{normzir}) reduces to
the familiar expression $S_E^{1/2}$ \cite{co85}.

\section{Related instantons}

The formalism used in the previous sections can be easily extended
to study a few more instantons which are closely related to the ones we
have seen so far, and which will be relevant for future discussion.

\subsection{Nucleation of topological defects during inflation}
\label{spontaneous}

We saw in Subsection \ref{instantonsub} that for $H\neq 0$ the action for
the instantons remains finite when the external field is switched off. This
corresponds to the spontaneous nucleation of membranes (or topological
defects), due to the gravitational field alone.

When the external field
$E_0$ is zero, there is no need to restrict ourselves to co-dimension 1.
Thus monopole pairs, strings and domain walls can spontaneously
nucleate in $d$-dimensional de Sitter space, with $d>N$.
The corresponding instantons are found by
intersecting the $d$-sphere with the necessary number of hyperplanes
through the origin:
\beq
\sum_{A=0}^dX^AX^A=H^{-2} \label{necessary}
\eeq
$$
X^i=0, \quad(i=N+1,...,n),
$$
which gives N-spheres of radius $H^{-1}$.
(here we use the lower case latin index $i$ for later notational
convenience).

When studying small perturbations around the instanton
for co-dimension larger than 1,
we will have more scalar fields `living' on the worldsheet,
one for each normal direction.
The generalization of (\ref{phinmu}) is
\beq
\delta x^{\mu}(\xi^a)=\mm^{-1/2}\sum_{i=N+1}^d\phi^{(i)}(\xi^a)n^{(i)\mu}.
\label{phinmui}
\eeq
Taking the normal vectors to be perpendicular to the
hyperplanes $X^i=0$, the effective action for $\phi^{(i)}$ is
\cite{baal91,gavi91,jemal2}
\beq
S^{(2)}[\phi]={1\over 2}\sum_{i=N+1}^d\int d^N\xi\sqrt{\gamma}[\phi^{(i)}
(-\Delta +M^2)\phi^{(i)}],\label{sphii}
\eeq
where $M^2$ is still given by (\ref{m2}) with $R_0=H^{-1}$, i.e.
$$
M^2=-NH^2.
$$
The fields can be expanded in terms of spherical harmonics on the
$N-$sphere, $\phi^{(i)}=\sum C^{(i)}_{LJ}\phi_{LJ}$,
and the normalization of the
zero modes can be worked out in exactly the same way as before. For an
infinitessimal rotation in the $(X^J, X^i)$ plane (where $J=0,...,N$,
$i=N+1,...,d$) of angle $d\alpha_J^{(i)}$ we have
\beq
dC^{(i)}_{1J}=H^{-1}d\alpha_J^{(i)}\left({\mm\sn(H^{-1})\over
N+1}\right)^{1/2}.\label{normi}
\eeq
Rotations of the $X^J$ among themselves or of the $X^i$ among themselves
leave the instanton invariant, so these will not correspond to zero modes.
Then, the total number of zero modes is $(N+1)(d-N)$.

In general, of the
$(N+1)(d-N)$ zero modes, $d$ will correspond to space-time translations and
the remaining $N(d-N-1)$ correspond to rotations in the spatial
orientation of the defect \cite{baal91}.
For instance, monopole pairs in 3+1
dimensional de Sitter space can nucleate with all possible orientations of
the relative position, and loops of string can nucleate with all possible
orientations of the plane of the loop.

\subsection{Particle at finite temperature in flat space}
\label{wrap}

For a massive particle at finite temperature the Euclidean action is
\beq
S_E=\mm\int_0^{\beta}(1+\dot{\vec x}^2(t_E))^{1/2} dt_E,
\label{hot}
\eeq
where $t_E$ is the Euclidean time and $\vec x(t_E)$ is the Euclidean
trajectory, which has to be periodic $\vec x(t_E)=\vec x(t_E+\beta)$.
Here $\beta$ is the inverse of the temperature.
The instantons for this system are straight lines
wrapping around the compact temporal dimension
\beq
\vec x(t_E)=\vec x_0=const. \label{lsy}
\eeq
The second
variation of the action is found by substituting
$\vec x(t_E)=\vec x_0+\mm^{-1/2}\vec \phi(t_E)$ in (\ref{hot}),
\beq
S^{(2)}_E={1\over 2}\int_0^{\beta}(\dot{\vec\phi})^2 dt_E. \label{vecphi}
\eeq
The fluctuations
$\vec\phi$ can be thought of as a set of massless scalar fields in 0+1
dimensions. From (\ref{vecphi}), $\vec\phi$ satisfies $\ddot{\vec\phi}=0$.
The zero mode solution $\vec\phi=const.$ amounts to a spatial translation
of the original solution $\vec x=\vec x_0$.

Expanding $\vec\phi$ in ``spherical harmonics'' on the 1-sphere
\beq
\vec\phi=\beta^{-1/2}\vec C_0
+\sqrt{2\over\beta}\sum_{L=1}^{\infty}\left\{\vec C_L
\cos\left[L{2\pi t_E\over \beta}\right]+\vec
D_L\sin\left[L{2\pi t_E\over\beta}\right]\right\},
\eeq
we see that the zero mode is in the $L=0$ sector. An infinitessimal
translation $d\vec x_0$ corresponds to
\beq
d\vec C_0=d\vec x_0 \sqrt{\mm\beta}.
\label{normb}
\eeq
The instanton (\ref{lsy}) and the previous equation will be used to
derive  the partition function for a gas of massive particles at finite
temperature.

\subsection{Pair production in more dimensions}
\label{schwpert}

In curved space, $E=const$ is a
solution of the homogeneous Maxwell's equations only in 1+1
dimensions. However, in flat space one can take $E=const$ in arbitrary
number of dimensions.
For completeness, we shall briefly consider the well known 3+1 dimensional
case in flat space.
The action for a particle in an external field is still given
by
\beq
S_E=\mm\oint  ds+e\oint A_{\mu}dx^{\mu},
\label{schw}
\eeq
where we are integrating over a closed worldline. If the
electric field $E$ is in the $x$ direction $\vec E=(E,0,0)$, we can take
$A_y=A_z=0$. It is clear from the previous sections that there will be an
instanton solution which is a circle of radius $R_0=\mm(eE)^{-1}$ in the
$(t_E,x)$ plane, where $t_E$ is Euclidean time. To find the second
variation of the action, we take a perturbation to the instanton of the
form (\ref{phinmui}). One of the normal vectors $n^{(r)}$ is chosen in the
outward radial direction from the origin of the $(t_E,x)$ plane. The
corresponding perturbation is denoted by $\phi_r(s)$, where $s\equiv R_0
\arctan(t_E/x)$ parametrizes the worldline. The second variation of $S_E$
with respect to $\phi_r$ is independent of the existence of the two
orthogonal directions $y$ and $z$, so from subsection $a$, the quadratic
action for $\phi_r$ is that of a scalar field of mass $M^2=-R_0^{-2}$ in
0+1 dimensions.

The other two normals are chosen in the $y$ and $z$ directions. The
associated perturbations $\phi_y(s)$ and $\phi_z(s)$ do not change the second
term in (\ref{schw}), since $A_y=A_z=0$. Therefore, from the previous
subsection these
perturbations will be massless fields. Putting it all together
\beq
S^{(2)}_E={1\over 2}\int_0^{2\pi R_0}
ds\left[\left(\dot\phi^2_r-R_0^{-2}\phi_r^2\right)+\dot\phi^2_y+\dot\phi_z^2
\right].
\label{pertschw}
\eeq
The zero modes for the massless fields are in the $L=0$
sector. They correspond to translations in the $y$ and $z$ directions and
can be normalized from (\ref{normb}), with $\beta=2\pi R_0$,
\beq
{dC_0^{(y)}\over dy}={dC_0^{(z)}\over dz}
=\sqrt{2\pi R_0\mm},\label{zirjul1}
\eeq
The zero modes
for $\phi_r$ are in the $L=1$ sector and they can be normalized from
(\ref{normzir})
\beq
{dC^{(r)}_{1t}\over dt_E}={dC^{(r)}_{1x}\over dx}
=\sqrt{\pi R_0\mm},\label{zirjul2}
\eeq
where we have replaced $H^{-1} d\alpha_J$ by the corresponding flat space
limit $dt_E$ or $dx$.

\section{Semiclassical partition function}

For systems at finite temperature, the lifetime of a metastable state
is related to the imaginary part of the free energy \cite{af81,co85} [see
Eq. (\ref{imf}) below]. The free energy is defined as
\beq
F\equiv -\beta^{-1}\ln Z, \label{free}
\eeq
where $\beta^{-1}$ is the temperature and $Z$ is the partition function
\beq
Z\equiv tr[e^{-\beta\hat H}],\label{Z}
\eeq
with $\hat H$ the Hamiltonian of the system.
The key to the semiclassical evaluation of $Z$ is to first express it as
a path integral. For instance, for the case of a single non-relativistic
particle in flat space moving in a potential $V(x)$, one has (see e.g.
ref. \cite{GPY82} for a nice discussion)
\beq
Z=\int_{x(0)=x(\beta)}\dd x(t) e^{-S_E},
\label{epi}
\eeq
where the integral is over all paths which are periodic in Euclidean time
with periodicity $\beta$, and
\beq
S_E=\int_0^{\beta} dt [{1\over 2}\dot x^2(t)+V(x(t))],
\eeq
is the Euclidean action.

As is well known \cite{giha77}, de Sitter space behaves in some respects
like a system at finite temperature $\beta^{-1}=H/2\pi$. One difference
with flat space is, however, that on the sphere all directions are compact.
Therefore, once in Euclidean there is no actual distinction between temporal
and spatial directions. In our case, the Euclidean action is given by
(\ref{mohi}) and the natural generalization of (\ref{epi}) is
\beq
Z=\int\dd\ws(\xi^a) e^{-S_E[\ws]},\label{epiws}
\eeq
where now the integral is taken over all closed worldsheets (or worldlines
for $N=1$).

In the semiclassical limit, $Z$ will be
dominated by the stationary points of $S_E$, and so it will be
a sum of contributions from
multi-instanton configurations $Z=Z_0+Z_1+...\ $. Here $Z_k$ is the
contribution of a configuration with $k$ widely separated instantons. This
configuration has action $k\bar S_E$, where $\bar S_E$ is the action for
one instanton. Hence $Z_k$ will have
the exponential dependence
\beq
Z_k\propto e^{-k \bar S_E}.
\eeq
To find the preexponential factor, one has to integrate over small
fluctuations around the stationary points. Eq.(\ref{epiws}) is
rather formal, because we have not specified the measure of integration
(this can actually be quite complicated, due to reparametrization
invariance.)
However, if all we are interested in are small fluctuations around the
instanton solutions, the integration over neighboring worldsheets amounts to
an ordinary path integral over the perturbation fields $\phi(\xi^a)$ that
we introduced in section \ref{perturbations}.
Therefore, we can write
\beq
Z=\sum_{k=0}^{\infty}{e^{-k\bar S_E}\over k!}\left(\int\dd \phi
e^{-S_E^{(2)}[\phi]}\right)^k,\label{yua}
\eeq
where $S_E^{(2)}$ is the second variation of the action, given
in (\ref{sphi}).
The sum is over multi-instanton configurations. The path integrals
over $\phi$ give the contribution of fluctuations around each one of
the spherical worldsheets. The $k!$ in the denominator can be understood as
follows \cite{co85}. When integrating over $\phi$ we are integrating also
over the zero modes. That means that we are integrating over all possible
locations of the instantons in Euclidean space. Since the instantons are
identical, we must divide by $k!$ to avoid overcounting.

Eq. (\ref{yua})
can be rewritten as
\beq
Z=e^{Z_1}, \label{gcz}
\eeq
where
\beq
Z_1\equiv e^{-\bar S_E}\int\dd\phi
\exp\left(\int\phi(\Delta+M^2)\phi\sqrt{\gamma}d^N\xi\right).\label{bnd}
\eeq
This is just the path integral for a free scalar field on a curved
background (the sphere), and can be calculated using well known
manipulations.

Following \cite{ha77}, the field is expanded in spherical harmonics
$\phi_{LJ}$ [see (\ref{harmonex})]. The integral over $\phi$ can then be
expressed in terms of the coefficients $C_{LJ}$
\beq
\dd \phi=\prod_{LJ} \mu {d C_{LJ}\over (2\pi)^{1/2}}.\label{mea}
\eeq
Note that $Z_1$ is dimensionless, whereas $C_{LJ}$ has dimensions of
$(mass)^{-1}$. To render $\dd \phi$ dimensionless one has to introduce the
parameter $\mu$ with dimensions of mass. Using (\ref{harmonex})
and (\ref{norma}) one has
\beq
Z_1=e^{-\bar S_E}\prod_{LJ}\int\mu {d C_{LJ}\over (2\pi)^{1/2}}
\exp\left({-{1\over 2}[\sum_{LJ}(M^2-\lambda_L)C_{LJ}^2]}\right),
\label{int}
\eeq
where $\lambda_L$ is given by (\ref{lambdal}). After Gaussian integration
one obtains
\beq
Z_1=e^{-\bar S_E}\prod_L[(\mu R_0)\Lambda_L^{-1/2}]^{g_L}
\equiv e^{-\bar S_E} (\det[(\mu R_0)^{-2}\hat O])^{-1/2},
\label{zdet}
\eeq
where $g_L$ is given by (\ref{gl}). Here we have introduced the
dimensionless operator $\hat O\equiv R_0^2(-\Delta+M^2)$, with eigenvalues
$\Lambda_L=R_0^2(M^2-\lambda_L),\ $[see (\ref{lambdal})].

Since the eigenvalues are known, the determinant can be calculated with
the $\zeta$-function regularization method. In terms of the generalized
$\zeta$-function,
\beq
\zeta (z)\equiv\sum_L g_L\Lambda_L^{-z},
\label{zeta}
\eeq
the determinant can be expressed as
\beq
\det[(\mu R_0)^{-2}\hat O]=(\mu R_0)^{-2\zeta(0)}
e^{-\zeta'(0)}.\label{detz}
\eeq
The values of $\zeta(0)$ and $\zeta'(0)$ for the operator $\hat O$ on the
$N$-sphere are calculated in the Appendix.
For $N=1$ we obtain $\zeta(0)=0$ and $\zeta'(0)=-2\ln[2\sinh(\pi R_0 M)]$, so
that
\beq
Z_1={e^{-\bar S_E}\over 2\sinh(\pi R_0 M)}.\quad (N=1)
\label{z1sh}
\eeq
Note that, since $\zeta(0)=0$, the dependence on the arbitrary
renormalization scale $\mu$ disappears.

For worldsheets of dimension $N>1$
it turns out that $\zeta(0)=0$ only for odd $N$ \cite{dona92}.
In general, for even $N$ the determinant will depend on the
renormalization scale $\mu$. As mentioned before, the calculation of $Z_1$
is the same as the calculation of the effective action for a free scalar
field in a curved spacetime of dimension $N$. The appearance of a
renormalization scale in that context is well known \cite{bida82,ha77} (in
particular, this scale is responsible for the so-called trace anomaly, the
quantum mechanical breakdown of conformal invariance). The difference
between even and odd dimensions can be understood from the fact that
in dimensional regularization the infinities come from the
poles of the gamma function $\Gamma[j-(N/2)]$, where $j$ is an integer.
As a result, for odd $N$ the effective action is finite after dimensional
regularization, and the usual $\log \mu$ terms do not appear.
In general,
for even $N$ we
have to live with the arbitrary scale $\mu$, unless $\zeta(0)$
happens to vanish accidentally for our particular worldsheet geometry and
mass of the scalar field $\phi$. It is well known that $\zeta(0)$ can be
computed in terms of geometrical invariants. For $N=2$ we have
\beq
\zeta(0)={1\over 4\pi}\int d^2\xi^a\sqrt{\gamma}\left[-m^2+\left({1\over
6}-\xi\right){\cal R}^{(2)}\right],
\label{zeta0}
\eeq
where $m^2+\xi {\cal R}^{(2)}$ is the quantity that we have denoted as
$M^2$ [see Eq. (\ref{m2comp})]. Notice that both in flat and de Sitter
backgrounds, the term involving
the external Ricci tensor in (\ref{m2comp}) is constant,
and can be included in $m^2$. We shall come back to the discussion of $\mu$
and its role in the context of renormalization in the next Section.

Our derivation of the semiclassical partition function has been rather
formal. As a check,
let us apply these ideas to a simple example where the result is known:
the case of free particles at finite temperature in flat space.
This will also illustrate the general procedure for dealing with the
zero modes.

For particles at finite temperature, the action for the instantons
(discussed in subsection \ref{wrap}) is equal to the mass $\mm$
of the particle
times the length of a worldline that wraps around the compact temporal
dimension $\bar S_E=\mm\beta$. The perturbation field $\phi$ is massless,
$M=0$, and so the expression (\ref{z1sh}) diverges because of the
vanishing denominator. This is to be expected because the operator $\hat O$
has a zero eigenvalue corresponding to the translational zero mode
discussed in section \ref{perturbations} (for simplicity we start by
considering only one spatial dimension transverse to the worldline, hence
only one field $\phi$ and one zero mode).

Noting that $(2\pi)^{-1/2}\int d
C_0 \exp(-M^2C_0^2/2)=M^{-1}$, the divergence at $M\to 0$ can be
avoided  by leaving the integral over $dC_0$ undone and
by excluding the factor $M^{-1}$
from the determinantal factor. Then we can write
\beq
dZ_1= (det '[(\mu R_0)^{-2}\hat O])^{-1/2} e^{-\mm\beta} {dC_0\over
(2\pi)^{1/2}},
\label{dz1}
\eeq
where
\beq
det '[(\mu R_0)^{-2}\hat O] \equiv \lim_{M\to 0}
{det[(\mu R_0)^{-2}\hat O] \over M^2}
=(2\pi R_0)^2 \label{detp}
\eeq
is the determinant without the zero eigenvalue.

Using (\ref{normb}) and $2\pi R_0=\beta$ we have
$$
dZ_1=dx_0 \left({\mm\over 2\pi\beta}\right)^{1/2}e^{-\mm\beta}.
$$
This can be cast in a more familiar form by setting $T\equiv\beta^{-1}$ and
increasing the number of transverse dimensions to three. Each transverse
dimension brings an additional power to the preexponential factor. Interpreting
$d^3x_0$ as the volume element we have
\beq
Z_1=V\left({\mm T\over 2\pi}\right)^{3/2} e^{-\mm/T}.\label{th}
\eeq
This is the correct expression for the ``one particle'' partition function
of an ideal gas with the Maxwell-Boltzmann distribution [note that the
instanton method is only valid when the exponent in (\ref{th}) is large, in
which case there is no difference between bosonic and fermionic
distributions].
The grand
canonical partition function is obtained, according to (\ref{gcz}), by
exponentiating this expression.

Actually, Eq. (\ref{gcz}) gives the partition function for the case of
vanishing chemical potential, something that we have tacitly assumed in our
derivation. The effect of a chemical potential $\tilde\mu$ is to replace
$Z_1$ in (\ref{gcz}) by $e^{\tilde\mu\beta}Z_1$. The number of particles
$\nn$ in the volume $V$ is then given by
$$
\nn=\beta^{-1}\left({\partial\ln Z
\over\partial\tilde\mu}\right)_{\beta,V}=
e^{\tilde\mu\beta}Z_1.
$$
For vanishing $\tilde\mu$, we have
\beq
\nn=Z_1.
\label{th2}
\eeq
Therefore $Z_1$ is also the equilibrium number of particles.

\section{Nucleation rates in flat space}
\label{prefactors}

In flat space and at sufficiently low temperatures, the decay rate of
a metastable state is given by \cite{af81}
\beq
\Gamma=2|Im\ F|=2\beta^{-1}|Im \ Z_1|.
\label{imf}
\eeq
(For temperatures $\beta^{-1}>R_0^{-1}$, this formula has to be
modified \cite{af81}. We defer the consideration of the high temperature
regime to a later paper.)
As shown in the previous section, the calculation of $Z_1$ reduces to the
calculation of a functional determinant. The fact that $F$ has an imaginary
part is due to the fact that the action $S^{(2)}$ in (\ref{sphi}) has a
negative mode, corresponding to $L=0$, so the determinant will be negative.
Upon taking the square root a factor of $i$ will emerge.

The membrane creation rate can be expressed as
\beq
d\Gamma={1\over 2}\times 2\beta^{-1}|det '[(\mu R_0)^{-2}\hat O]|^{-1/2}
e^{-\bar S_E}|J|d V dt_E,
\label{wf12}
\eeq
where the Jacobian is given by
\beq
|J|={\prod_{J=0}^{d-1}(2\pi)^{-1/2}dC_{1J}\over dVdt_E}.
\label{jacobian}
\eeq
This equation follows from (\ref{imf}) and (\ref{zdet}). As before [see
(\ref{detp})] the prime in the determinant means that the
$(N+1)$ zero modes (which
are now in the $L=1$ sector) are omitted, because the integration over
$dC_{1J}$ is left undone
\beq
det '[(\mu R_0)^{-2}\hat O]\equiv\lim_{M^2\to -N R_0^2}
{(\mu R_0)^{-2\zeta(0)} e^{-\zeta'(0)}\over (M^2+N R_0^2)^{N+1}}.
\label{ndetp}
\eeq
The Jacobian, which is
needed to change variables from $dC_{1J}$ to the
Euclidean space-time volume element $dV\ dt_E$, can be read off from
(\ref{normzir}) [for obvious reasons, in the limit $H\to 0$ we replace
$H^{-1}d\alpha_J$ by $dt_E$ (for $J=0$) or by $d\vec x$ (for $J=1,...,d-1)$],
$$
|J|=\left({\mm\sn(R_0)\over 2\pi(N+1)}\right)^{N+1\over 2}.
$$
The overall factor of $1/2$ in the r.h.s. of (\ref{wf12}) is explained in
Refs. \cite{la67,co85}. It arises because the free energy of an unstable
state can only be defined by analytic continuation from a Hamiltonian in
which the same state is stable. As a result, the contour
of integration over the negative mode $dC_0$ has to be deformed into the
complex plane in such a way that only half of the saddle point contributes
to $Im\ F$. We should note, however, that in the derivation of (\ref{imf})
given in \cite{af81}, these considerations are not really relevant; and
$2|Im\ F|$ is essentially a convention to denote the r.h.s. of (\ref{wf12}).

Integration over Euclidean time $t_E$ cancels the factor of $\beta^{-1}$,
and we are left with the rate per unit volume
\beq
{d{\cal N}\over dt dV}=
{\Gamma\over V}=\left({\mm\sn(R_0)\over 2\pi (N+1)}\right)^{N+1\over 2}
|det '[(\mu R_0)^{-2}\hat O]|^{-1/2} e^{-\bar S_E},
\label{rate}
\eeq
where ${\cal N}$ is the number of membranes created.
Let us now evaluate this rate for space-times of different dimensionalities.

\subsection{Pair creation in 1+1 dimensions}

{}From (\ref{ndetp}) and (\ref{zpn1}) the primed determinant is
$$
\lim_{M^2\to-R_0^2}{4\sinh^2(\pi R_0M)\over(M^2+R_0^{-2})^2}=-\pi^2 R_0^4,
$$
hence, from (\ref{rate}) the creation rate per unit length is
\beq
{\Gamma\over L}={\mm\over 2\pi R_0}e^{-\bar S_E}.
\label{stone}
\eeq
Taking $R_0=\mm/eE_0$, we have
$$
{\Gamma\over L}={eE_0\over 2\pi}\exp\left(-{\pi\mm^2\over eE_0}\right).
$$
This can be compared with the results of Stone \cite{st76},
who computed the vacuum decay rate in the
in the sine-Gordon theory with non-degenerate vacua. His one loop
result was
$$
{\Gamma\over L}={eE_0\over 2\pi} |ln(1-e^{-{\pi\mm^2\over eE_0}})|,
$$
where $\mm$ is the mass of the kink and $eE_0$ is the vacuum energy
density gap between neighboring vacua. As expected, the instanton
calculation gives a good approximation when the Euclidean action
is large $\bar S_E>>1$.

Eq. (\ref{stone}) also gives the rate at which a metastable cosmic string
will break up by nucleating pairs of monopoles \cite{prvi92}. In that
case $eE_0$ should be replaced by the string tension and $\mm$ by the mass
of the monopoles.

\subsection{Pair creation in 3+1 dimensions}

As explained in subsection \ref{schwpert}, in 3+1 dimensions in addition to
the radial perturbations of mass $M^2=-R_0^2$, we have the
perturbations $\phi_y$ and $\phi_z$ which are transverse to the plane of
the instanton. These behave like massless fields $M^2=0$. Each field
contributes its own determinantal factor, which for $\phi_y$ and $\phi_z$
is given by  (\ref{detp}). Also, from (\ref{zirjul1}), each contributes a
factor $(R_0\mm)^{1/2}$ to the Jacobian. With this, Eq. (\ref{stone}) is
modified into
\beq
{\Gamma\over V}={(eE_0)^2\over 8\pi^3}e^{-{\pi\mm^2\over eE_0}}.
\label{schwinger}
\eeq
This coincides with the rate of production of charged bosons in scalar
electrodynamics
\beq
{\Gamma\over V}={(eE_0)^2\over 8\pi^3}\sum_{n=1}^{\infty}{(-1)^{n+1}\over
n^2}e^{-n{\pi\mm^2\over eE_0}}[1+O(e^2)],
\label{sqed}
\eeq
in the limit $S_E>>1$. (For a more thorough account of
monopole production by a magnetic field,
and pair production in the strong coupling regime see
Refs. \cite{afma82,AAM82}).

\subsection{String creation in 2+1 dimensions}
\label{gunthersub}
{}From (\ref{rate}) and (\ref{wak}),
\beq
{\Gamma\over V}=\left({\mm{\cal S}_2(R_0)\over 6\pi}\right)^{3/2}
(\mu R_0)^{7/3} R_0^{-3} e^{-\bar S_E}.
\label{gunther}
\eeq
Note that, since $\zeta(0)\neq 0$, the determinant depends explicitly
on the renormalization scale $\mu$ [the exponent $7/3$ can also be derived
from (\ref{zeta0}).] Because of the arbitrariness in $\mu$ we cannot give
an absolute estimate of the nucleation rate the way we do for N=1 or N=3.
However, since $\mu$ is a constant, we can still predict how the rate
changes when we change the external field $E_0$ (that is, when we change
$R_0$.) It is seen that the dependence of the prefactor on $R_0$ is more
complicated than what one would have guessed form dimensional analysis.

So far, by using the $\zeta$-function method, we have avoided the question
of ultraviolet divergences, since they are automatically removed by
analytic continuation \cite{bida82} (see also Ref.\cite{nami} for a recent
discussion, and references therein).
However, we should recall that the functional
determinants contain such divergences, and that these can be elliminated by
suitable counterterms in the action. For $N=2$, all divergences can be
removed by counterterms of the form \cite{bida82}
$$
c\int d^2\xi\sqrt{\gamma}+ d\int d^2\xi \sqrt{\gamma}R.
$$
The first term is a contribution to the membrane tension. The second is a
topological invariant which does not contribute to the equations of motion.
Note, form (\ref{detz}) and (\ref{zeta0}),
that a rescaling of the arbitrary parameter $\mu$ can be reabsorbed in
a redefinition of $c$ and $d$. Then, we can elliminate the
renormalized $c$ by rescaling $\mu$, and the renormalized $d$ by shifting
the string tension $\mm$.

In Ref. \cite{GNW80} a different
approach was followed in which the product
over all eigenvalues was cut off at some
physical scale $\mu$. This method
also produced the factor $(\mu R_0)^{7/3}$, which was refered to as the
``universal term'' \cite{GNW80,af79}.
The introduction of a physical cut-off is not unreasonable
when dealing with an effective theory.
In deriving the action for the perturbations we
have only expanded to second order in $\phi$. This is justified at low
momenta $L$, but in the ultraviolet limit, higher order terms become
important. To realize that this is so, let us take the case of a flat
membrane and $E_0=0$ (this is actually quite representative of the
general case). The exact Nambu action is then
$$
S_E=\mm\int d^2\xi \sqrt{1+\mm^{-1}(\partial_a\phi\partial^a\phi)}.
$$
When $\partial_a \phi$ is of order $\mm^{1/2}$ higher order terms become
important and the quadratic approximation fails. This suggests taking
$\mu\sim\mm^{1/2}$ as a cut-off,
which is essentially what was done in \cite{GNW80}.

\subsection{Membrane creation in 3+1 dimensions}

Using (\ref{rate}) and (\ref{wek}) we have
\beq
{\Gamma\over V}=\left({\mm{\cal S}_3(R_0)\over 8\pi}\right)^2
{4R_0^{-4}\over\pi^2}
e^{\zeta'_R(-2)} e^{-\bar S_E}.
\label{affleck}
\eeq
This can be compared with the results of Affleck \cite{af79}. He studied
the decay rate of false vacuum in the theory of a scalar field with a
symmetry breaking potential and nondegenerate vacua,
in the limit in which the thickness of the wall separating the true
from the false vacuum is much smaller than
the radius of the bubble at nucleation.

The first factor in the r.h.s. of (\ref{affleck}) is the Jacobian $|J|$
that comes from the normalization of the zero modes. The rest of the
pre-exponential factor
coincides with what Affleck calls the ``universal terms''.
These are due to fluctuations of the worldsheet and therefore are
independent of the details of the field theoretic model. That is, they
are independent of the
internal structure of the ``kink'' or domain wall.
The evaluaton of quantum corrections to the effective action due to finite
thickness of the wall is in itself an interesting subject, and the
corrections can be important in realistic theories
\cite{af79,afma82,AAM82,FSM84}. However, since these corrections are not
essential to the physics of nucleation here we shall concentrate on
infinitely thin membranes, without internal structure. We will only mention
that the fluctuations around thick walls can be treated very naturally
within the formalism described in this paper. In addition to the tachyonic
field of perturbations that we have considered, there will be an infinite
tower of massive fields $\phi_q$ `living' in the N-sphere
\cite{vavi91,gavi92}, each one contributing its own determinantal
prefactor. We plan to return to this question in a later paper where the
thermal properties of the fields $\phi_q$ (with temperature $2\pi R_0$) are
emphasized.\cite{GVV93}

\section{Nucleation rates in curved space}

Although the calculation of determinants for the instantons in de Sitter
space offers no special problems (we only
need to substitute the appropriate values of $R_0$ in the expressions found
in the Appendix); the definition of a nucleation
rate in de Sitter is a more subtle issue
which has often been eluded in the literature.
One possibility would be
to simply use Eq. (\ref{rate}), interpreting $dV$ as the physical volume
element at the time of nucleation $dV_0$
\beq
d\Gamma=|\lambda|dV_0,\label{rama}
\eeq
where $|\lambda|$ is defined as the r.h.s. of (\ref{rate}).

Although we believe that this expression is correct
(when properly interpreted) it clearly needs further justification.
First of all, the physical volume element, $dV_0$, is proportional to a
power of the scale factor $e^{Ht_0}$ at the time of nucleation. If Eq.
(\ref{rate}) was derived from a purely Euclidean calculation, how does the
exponential of a Lorentzian time find its way into the r.h.s. of
(\ref{rama})? Also, as pointed out in \cite{baal91}, the time of nucleation
in de Sitter space is a somewhat ambiguous concept when the size of the
instantons is comparable to the horizon, and in principle it is
not clear what time one should use in $dV_0$.

These difficulties prompt us to search for an alternative way of
calculating nucleation rates in de Sitter space, which does not take
(\ref{imf}) as the starting point. Recalling that de Sitter space behaves
in some respects like a thermodynamical system, one can try to estimate
directly the equilibrium distribution of membranes. For this one can use
Eq. (\ref{th2}), with ${\cal N}$ the number of membranes and $Z_1$ given by
(\ref{bnd}). It is clear that
$$
d\nn=(det'[(\mu R_0)^{-2}\hat O])^{-1/2} e^{-\bar S_E}\prod_{J=0}^{d-1}
(2\pi)^{-1/2}dC_{1J}.
$$
Using (\ref{normzir}), we have
$$
d{\nn}=\lambda H^{-d}\prod_{J=0}^{d-1}d\alpha_J
$$
where
\beq
\lambda=\left({\mm\sn(R_0)\over 2\pi(N+1)}\right)^{N+1\over 2}
(det'[(\mu R_0)^{-2}\hat O])^{-1/2} e^{-\bar S_E}.
\eeq
Eq. (\ref{normzir}) is valid only for infinitesimal rotations, and
in that case $\prod d\alpha_J$ can be identified with the differential
solid angle on the d-sphere, $d\Omega$,
within which the center of the instanton worldsheet is to be found,
\beq
d \nn=\lambda H^{-d}d\Omega. \label{omega}
\eeq
One can interpret this equation as the ``equilibrium distribution of
instantons'' in Euclidean space.

Of course, an equilibrium distribution of instantons is not a measurable
object. However, it can be given meaning by conjecturing that the
equilibrium distribution of membranes in the Lorentzian section is given by
the analytic continuation of the previous object to real time. This
prescription is just heuristic, and we do not know how to justify it
further, except by saying that it reduces to (\ref{rate}) in flat space and
that it is quite natural from the mathematical point of view.

To see how the analytic continuation is done, let us consider, for
simplicity, the 1+1 dimensional case (higher dimensional cases are
completely analogous). Then (\ref{omega}) reads
$d\nn=\lambda H^{-2}\cos\alpha_E d\alpha_E d\beta$, where $\alpha_E$ and
$\beta$ are polar and azimuthal angles on the 2-sphere. Upon analytic
continuation $\alpha_E=i\alpha$ [see (\ref{iguana})],
$$
d\nn=|\lambda|H^{-2}\cosh\alpha d\alpha d\beta.
$$
Note that the factor of $i$ from $d\alpha_E$ cancels the imaginary factor
from the square root of
the determinant in $\lambda$, so that $d\nn$ is actually real. Using
Eqs. (\ref{x0w}) and (\ref{t0w}), one can change variables from
$(\alpha,\beta)$ to $(x_0,t_0)$. The Jacobian is
$$
\left|{\partial(t_0,x_0)\over\partial(\alpha,\beta)}\right|=
H^{-2}\cosh\alpha e^{-Ht_0},
$$
and therefore
\beq
d\nn=|\lambda|e^{Ht_0}dx_0dt_0.\label{rama11}
\eeq
Generalizing to spacetimes of arbitrary dimension we have
\beq
d\nn=|\lambda|e^{NHt_0}d\vec x_0 dt_0,\label{rama12}
\eeq
which is very similar to (\ref{rama}), but now all ambiguities in $dV_0$
have been resolved.

An equation of the same form as the previous
one was given in \cite{baal91}, based
on kinematical considerations. However, the parameter $\lambda$ was left
unspecified. Like in \cite{baal91}, Eq. (\ref{rama12}) is a
distribution in the space of parameters $x_0$ and $t_0$, and it is
therefore independent of the time of observation.

Following \cite{baal91}, we can use (\ref{genwl}) to express
$t_0$ in terms of the physical radius $R$, and thus find the size
distribution of membranes (or bubbles),
$$
{d\nn\over dV_{phys}}={|\lambda|\over H^d}
{R(R^2-R_0^2)^{-1/2}\over\lc\omega_0+(R^2-R_0^2)^{1/2}\rc^d}dR,
\label{bubbles}
$$
where $dV_{phys}=\exp(NHt)d^N\vec x_0$. Notice that the
distribution diverges at the lower end
$R\to H^{-1}$ when $\omega_0\leq 0$. This
divergence was interpreted in \cite{baal91}. For large radii, one finds the
scale invariant distribution
\beq
{d\nn\over dV_{phys}}\approx{|\lambda|\over H^d}{dR\over
R^d},\label{scaleinv}
\eeq
which depends on the external field $E_0$ and membrane tension only through
the coefficient $\lambda$. The fact that the size distributions are time
independent is easy to understand. As the bubbles are created, they are
stretched and diluted by the inflationary expansion, giving rise to a
stationary distribution of sizes \cite{baal91}.

The ``nucleation rates'' $|\lambda|$ for $d=2,3$ and $4$ can be read off
from the r.h.s. of (\ref{stone}), (\ref{gunther}) and (\ref{affleck})
respectively, where $R_0$ is given by (\ref{r0}) and $\bar S_E$ by
(\ref{se1}-\ref{se3}). This covers the case of co-dimension one, $d=N+1$.
For completeness, we shall also consider the case of strings and monopole
pairs spontaneously nucleating in 3+1 dimensions.

For the case of strings, the co-dimension is 2, and so there will be two
independent perturbation fields $\phi^{(i)}$ and two determinantal
prefactors. From (\ref{gunther}) it is clear that $\lambda$ will
contain one factor of $\mm^3$ and a
factor of $\mu^{14/3}$, where $\mu$ is the renormalization scale.
Taking $R_0=H^{-1}$, the
rest will be a numerical factor
(which  can be absorbed in $\mu$), times the appropriate power of $H$
necessary to give $\lambda$ the dimensions of $(mass)^4$:
$$
|\lambda|=\mu^{14/3}\mm^3 H^{-11/3}e^{-\bar S_E}.
$$
Because of the arbitrary renormalization scale, we cannot obtain an
absolute estimate for $|\lambda|$, but only its dependence on the expansion
rate $H$. Like in subsection (\ref{gunthersub}), to obtain a crude
absolute estimate one can set $\mu\sim\mm^{1/2}$, which gives
\beq
|\lambda|\sim \left({\mm\over H}\right)^{23/3}H^4\exp(-4\pi\mm H^{-2}).
\label{string}
\eeq
For $\mm>>H$ this can be considerably larger than the naive dimensional
estimate $|\lambda|\sim H^4\exp(-\bar S_E)$ mentioned in \cite{baal91}, or
even the more sophisticated $|\lambda|\sim \mm^3 H\exp(-\bar S_E)$.
Unfortunately, the existence of an arbitrary renormalization scale leaves
us quite uncertain as to the overall normalization of (\ref{string}).

Let us now consider the case of pairs spontaneously nucleating in $3+1$
dimensions. The co-dimension is 3 and so there will be 3 perturbation
fields $\phi^{(i)}$, each one of them with mass $M^2=-H^{-2}$,
contributing to the determinantal prefactor. Each field has 2 zero modes
[in the $L=1$ sector, see (\ref{normi})], which makes a total of 6. Four of
them correspond to space-time translations and two of them to changes in
the orientation of the monopole pair in three dimensional space. Thus we
have
$$
|\lambda|=\left({\mm H\over 2\pi}\right)^3\left({4\pi\over H^2}\right)
e^{-\bar S_E}.
$$
The factor of $(\mm H/2\pi)^3$ comes from taking the third power of the
prefactor in (\ref{stone}), with $R_0=H^{-1}$. The factor $4\pi H^{-2}$
is a correction due to the fact that two of the zero modes represent
changes in the angular orientation of the pair, rather than translations.
For each angular variable, the Jacobian (\ref{jacobian}) has an extra power
of $H$ in the denominator [see (\ref{normi})]. Integration over all
possible orientations gives the factor of $4\pi$. To summarize,
\beq
|\lambda|={H\mm^3\over 2\pi^2}\exp\lp-{2\pi\mm\over H}\rp,
\quad (d=3+1)\label{monopole}
\eeq
where $\mm$ is the mass of the monopole.

For the case of pair creation a `size' distribution such as (\ref{scaleinv})
is not very useful. Instead, it is more convenient to find the momentum
distribution of particles. Let us find the conserved momentum as a function
of the time of nucleation $t_0$.
In 1+1 de Sitter space, the vector potential
\beq
A_{\mu}=-H^{-1}E_0e^{Ht}\delta_{\mu x}
\label{vector}
\eeq
represents a constant electric field (note that
$F_{\mu\nu}F^{\mu\nu}=2E_0^2$). With this, the action for the point
particle coupled to $A_{\mu}$ reads
\beq
S=-\mm\int(\dot t-\dot x^2e^{2Ht})^{1/2} d\tau-{eE_0\over H}\int e^{Ht}\dot
x d\tau,
\label{cona}
\eeq
where $\tau$ is the proper time and a dot denotes derivative with respect
to $\tau$. Since the Lagrangian does not depend on $x$, the momentum
$$
k\equiv{\partial L\over \partial \dot x}=\mm\dot x e^{2Ht}-eE_0e^{Ht},
$$
is conserved. Setting $x_0=0$ in (\ref{genwl}) we have
$$
{dx\over dt}={H\omega_0e^{-H(t_0+t)}-e^{-2Ht}\over Hx}
$$
and from $\dot t^2-\dot x^2e^{2Ht}=1$, it follows that
$
\dot t={e^{Ht_0}H|x|(1-H^2\omega_0^2})^{-1/2}.
$
Therefore,
\beq
k=-{\mm\over HR_0}e^{Ht_0} sign(x),
\label{momentum}
\eeq
where we have taken into account that the particle to the left
of the inside region has charge of opposite sign.

Using this equation in (\ref{rama11}), with $|\lambda|$ given by the right
hand side of (\ref{stone}), we have
\beq
{d\nn\over dx_0}=e^{-\bar S_E}{dk\over 2\pi}. \quad (d=1+1)
\label{compare2}
\eeq
In 3+1 dimensions (without electric field), using (\ref{rama12}),
(\ref{monopole}) and (\ref{momentum}), we have
\beq
{d\nn \over d^3x_0}=
\exp\lp-{2\pi\mm\over H}\rp{k^2dk\over 2\pi^2}.\quad (d=3+1)
\label{compare4}
\eeq
Note that the
momentum distributions are flat, as expected on the grounds of scale
invariance.

At any given time $t$ we only need to integrate up to a cut-off
momentum
$$
|k|\sim{\mm\over HR_0}e^{Ht},
$$
since particles with higher value of the coordinate momentum have not been
created yet [see (\ref{momentum})]. Then
$$
n\sim {1\over 2\pi}{\mm\over HR_0}e^{-\bar S_E}, \quad (d=1+1)
$$
and
$$
n\sim {\mm^3\over 6\pi^2} e^{-{2\pi\mm\over H}}.\quad (d=3+1),
$$
where $n$ is the number density of particles per unit physical volume.
As noted in \cite{baal91}, for the case of vanishing electric field the
distribution of particles contains a Boltzmann factor $\exp(-\mm/T)$ where
$T=H/2\pi$ is the Gibbons-Hawking temperature \cite{giha77}.

\section{Charged scalar field in 1+1 dimensional de Sitter space}

The Klein-Gordon equation for a charged scalar field $\varphi$ coupled to
an external electromagnetic field $A_{\mu}$ is
\beq
-g^{\mu\nu}(\nabla_{\mu}-ieA_{\mu})(\nabla_{\nu}-ieA_{\nu})
\varphi+\mm^2\varphi=0.
\label{kg}
\eeq
As we shall see below, for a constant electric field in 1+1 de Sitter this
equation can be solved in terms of special functions. Then the problem of
particle creation is amenable for calculation using Bogolubov
transformations (see e.g. \cite{bida82}).

To apply this method it is necessary to specify an `in' state and an `out'
vacuum. The `in' state, $|in>$, is the physical quantum state of our system,
fixed by initial conditions. The `out' vacuum, $|0>_{out}$, is a different
quantum state in the Hilbert space, whose choice amounts to a definition of
particles at late times. Whether or not one can unambiguously specify
$|0>_{out}$ depends on whether or not it is physically reasonable to define
particles at late times. One way to guarantee a reasonable definition is to
switch off the gravitational and the electric fields at late times
(although this may not be necessary).

To illustrate the procedure, consider the case of vanishing electric field
first \cite{buda78}. In terms of the conformal time $\eta\equiv
-H^{-1}e^{-Ht}$, and with $\varphi=\varphi_k(\eta)e^{ikx}$, Eq. (\ref{kg})
reads
\beq
\vpk''+\left(k^2+{\mm^2\over H^2\eta^2}\right)\vpk=0,\label{buda}
\eeq
where a prime denotes derivative with respect to $\eta$.
This equation is symmetric in $k$, and following the usual convention, we
take $k>0$
(For $E_0=0$, the results for $k<0$ are the same).
Eq. (\ref{buda}) has the general solution
\beq
\vpk(\eta)=\left({\eta\over 8}\right)^{1/2}
\left[A_k\hnd(k\eta)+B_k\hnu(k\eta)\right],
\label{combo}
\eeq
where
$$\nu=\left({1\over 4}-{\mm^2\over H^2}\right)^{1/2}$$
and $H_{\nu}^{(i)}$ are the Hankel functions. For the `in' state, we shall
take the Bunch-Davies vacuum \cite{buda78}. This is characterized by
positive frequency modes of the form (\ref{combo}) with $A_k=1$, $B_k=0$
\beq
\vpkinp(\eta)=\left({\eta\over 8}\right)^{1/2}\hnd(k\eta).\label{vpkinp}
\eeq
The choice of this vacuum as the physical `in' state
can be motivated from many different points of
view, and it is a clear favourite in studies of inflation. In the open
coordinate system that we are using, this is the only truly de Sitter
invariant vacuum \cite{FHJ87}. The two-point function in this state
coincides with the Euclidean two-point function \cite{ha84}, which has
the important property of having the Hadamard form (roughly speaking, this
means that it has similar ultraviolet behaviour as the two-point function
in flat space). Also, it is believed that if the Universe nucleated from
`nothing' into a de Sitter phase, then this is the quantum state that the
fields would be in after nucleation \cite{haha85,vi88}.

To define the `out' vacuum it is convenient to write down the equation for
the scalar field in terms of the cosmological time $t$,
\beq
\ddot\vpk+H\dot\vpk+\left(\mm^2+{k^2\over
a^2(t)}\right)\vpk=0,\label{cosmolt}
\eeq
where a dot denotes $d/dt$ and $a(t)=\exp(Ht)$. Introducing
$\vpk=a^{-1/2}\psi_k$ we have
\beq
\ddot\psi_k+\left[\mm^2-{H^2\over 4}+{k^2\over a^2}\right]\psi_k=0.
\label{psicho}
\eeq
Let us take $\mm>>H$ and define $t_k$ as the time when the physical
wavelength $a(t)k^{-1}$ is equal to the particle's Compton
wavelength
\beq
ke^{-Ht_k}=\mm.
\eeq
For $t>>t_k$ the $k^2 a^{-2}$ term is negligible compared to $\mm^2$. Then
we will have approximate solutions of the form
$$
\vpk\approx a^{-1/2}\left[C_ke^{-iwt}+D_ke^{+iwt}\right]
$$
where
\beq
w=\lp \mm^2-{1\over 4}H^2 \rp^{1/2}.\label{omega2}
\eeq
Since $\mm>>H$, the exponentials oscillate very fast compared to the rate
at which $a^{1/2}$ changes, and so we will have an approximate definition
of positive and negative frequency `out' modes. For $t>>t_k$,
\beq
\vpkoutpm\propto a^{-1/2}e^{\mp iwt}.\label{particles}
\eeq
This definition is not very natural for $\mm^2<<H^2$, so we shall not
be considering this limit.

A purist would object that we cannot define particles unless the expansion
of the Universe is switched off. Consider then a FRW model in which the
expansion rate $\hh\equiv \dot a/a$ is time dependent. Then $\psi_k$
satisfies the equation
\beq
\ddot\psi_k+\lc\mm^2-{\hh^2\over 4}+{k^2\over a^2}-{1\over 2}\dot \hh\rc
\psi_k=0.\label{tempo}
\eeq
After a sufficiently long period of inflation, $a\propto\exp(Ht)$, the
expansion is adiabatically switched off starting at time $t_s$, in such a
way that $\dot \hh<<\mm^2$, until $a(t)$ reaches a constant value. The
space-time is Minkowski in the asymptotic  future. It is clear from
(\ref{tempo}) that if $t_s>t_k$, the mixing between positive and negative
frequency modes will be negligible during the period in which the expansion
is being switched off. This means that for modes such that $t_s>t_k$ the
number of particles that is calculated using the definition
(\ref{particles}) for positive and negative frequency modes is the same
that the number of particles
that would be found in the `out' Minkowski region.

The `in' positive frequency mode can be expressed as a linear combination
of the `out' positive and negative frequency modes
\beq
\vpkinp=\alpha_k\vpkoutp+\beta_k\vpkoutm.
\label{bogolubov}
\eeq
The coefficients $\alpha_k$ and $\beta_k$ are the so-called Bogoliubov
coefficients.
Using the asymptotic expression for the Hankel functions at late
cosmological times $t\to \infty$ (i.e. $\eta\to 0$)
$$
\vpkinp(k\eta)=-\lp{\eta\over 8}\rp^{1/2}
{i\over \nu\pi}\lc\lp{|k\eta|\over
2}\rp^{\nu}\Gamma(1-\nu)e^{-i{\pi\nu\over 2}}-
\lp{|k\eta|\over 2}\rp^{-\nu}\Gamma(1+\nu)e^{+i{\pi\nu\over
2}}\rc \times
$$
$$
[1+O(k\eta)],
$$
and the relation
\beq
|\alpha_k|^2-|\beta_k|^2=1,\label{normbog}
\eeq
one readily finds
\beq
|\beta_k|^2=[\exp(2\pi wH^{-1})-1]^{-1},\label{bose}
\eeq
where $w$ is given by (\ref{omega2}). This equation was found e.g. in Refs.
\cite{reta88,mo85} (see also references therein), using a different
coordinate system.

The number density of particles per unit coordinate
volume is then
\beq
{d\nn\over dx}=|\beta_k|^2{dk\over 2\pi}
\approx e^{-2\pi\mm H^{-1}}{dk\over 2\pi},
\quad (d=1+1),\label{compared2}
\eeq
where we have used $\mm>>H$. The same calculation can be done in $d$
spacetime dimensions. The only difference in the final result is that
$$
w=\lp\mm^2-{(d-1)^2\over 4}h^2\rp^{1/2}
$$
in Eq. (\ref{bose}).
Using $\mm^2>>H^2$ we have
\beq
{d\nn\over d^3\vec x}=
|\beta_k|^2{d^3 k\over (2\pi)^3}\approx
\exp\lp-{2\pi\mm\over H}\rp{k^2dk\over 2\pi^2}.\quad (d=3+1)
\label{compared4}
\eeq
Comparing (\ref{compared2}) and (\ref{compared4}) with (\ref{compare2}) and
(\ref{compare4}) we find complete agreement, not only in the exponential
behaviour, but also in the the preexponential factor.

Let us now consider the case of non-vanishing electric field in 1+1
dimensions. In conformal time, the Klein Gordon equation with $A_{\mu}$
given by (\ref{vector}) reads (note that $\nabla_{\mu}A^{\mu}=0$)
\beq
H^2\eta^2\vpk''+\lp k^2H^2\eta^2-2eE_0k\eta+\mm^2+{e^2E_0^2\over H^2}\rp
\vpk=0.\label{whiteq}
\eeq
This is the Whittaker equation, having the general solution
\beq
\vpk=A_kW_{\lambda,\sigma}(2ik\eta)+B_kW_{-\lambda,\sigma}(-2ik\eta),
\label{whitgen}
\eeq
where $W_{\lambda,\sigma}$ are the Whittaker functions, with
$\lambda=+ieE_0H^{-2}$ and
\beq
\sigma=\lp {1\over 4}-{\mm^2\over H^2}-{e^2E_0^2\over H^4}\rp^{1/2}.
\label{sigmain}
\eeq
At early times ($\eta\to-\infty$), Eq. (\ref{whiteq}) is very similar to
(\ref{buda}), and both reduce to
$$
\vpk''+k^2\vpk=0.
$$
Using the asymptotic expression for $\hnd(k\eta)$ at large $\eta$, the
`in' positive frequency mode in the Bunch-Davies vacuum has the form
$\vpkinp\sim(4\pi k)^{-1/2}\exp(-ik\eta)$, and so it is positive frequency
with respect to the conformal time. With the electric field switched on, we
would like to choose a positive frequency mode which has similar behaviour
at $\eta\to- \infty$. Noting that
$$
W_{\lambda,\sigma}(2ik\eta)\sim e^{-ik\eta}(ik\eta)^{\lambda},
$$
and since $W_{\lambda,\sigma}^*(z)=W_{-\lambda,\sigma}(-z)$,
it is clear that we have to take $B_k=0$ in (\ref{whitgen}),
\beq
\vpkinp=(4\pi k)^{-1/2}W_{\lambda,\sigma}(2ik\eta),\label{had}
\eeq
where the normalization is due to the usual Wronskian condition.

It is quite straightforward to show that the state defined by the modes
(\ref{had}) is a Hadamard vacuum. For this, one simply writes the two point
function as a sum over the modes (\ref{had}). Similarly, one can write the
Bunch-Davies two-point function as a sum over the modes (\ref{vpkinp}).
Because of the similarity in the ultraviolet behavior of both sets of
modes, it is easy to show that the difference between both two-point
functions is finite in the limit of coincident points, which simply means
that both two-point functions have the same singularity structure.

To define particles at late times ($\eta\to 0$) we
observe that, in this limit, Eqs. (\ref{buda}) and (\ref{whiteq}) are again
very similar, so the procedure that worked there will work here too. In
terms of cosmological time $t$ and the variable $\psi_k=a^{1/2}\vpk$ we have
\beq
\ddot\psi_k+\lp\mm^2+{e^2E_0^2\over H^2}-{H^2\over 4}+{k^2\over
a^2}-{2eE_0\over H}{k\over a}\rp\psi_k=0.
\eeq
Let us denote by $t_k$ the time at which the physical wavelength of the mode
is equal to the effective Compton wavelength
\beq
k e^{-Ht_k}=\lp\mm^2+{e^2E_0^2\over H^2}\rp^{1/2}.
\label{timek}
\eeq
Provided that
\beq
w\equiv \lp\mm^2+{e^2E_0^2\over H^2}-{H^2\over 4}\rp^{1/2}>>H,
\label{omegae}
\eeq
it is clear that for $t>>t_k$
we can define particles in the mode $k$, using as positive and negative
frequency modes the solutions
\beq
\vpkoutpm\propto a^{-1/2}e^{\mp iwt}.\label{130}
\eeq
As before, these definitions are not meaningful for $w<<H$, and we shall
not consider this limit.

Let us introduce the new Whittaker function $M_{\lambda,\sigma}$,
\beq
M_{\lambda,\sigma}(z)\equiv
\Gamma(2\sigma+1)e^{i\pi\lambda}
\lc e^{-i\pi(\sigma+{1\over 2})}
{W_{\lambda,\sigma}(z)\over \Gamma(\sigma+\lambda+{1\over 2})}+
{W_{-\lambda,\sigma}(-z)\over \Gamma(\sigma-\lambda+{1\over 2})}\rc.
\label{grry}
\eeq
$$
\lp -{3\over 2}\pi<\arg z<{1\over 2}\pi\rp
$$
For small $z=2ik\eta$ we have \cite{grry}
$$
M_{\lambda,\sigma}=z^{{1\over 2}+\sigma}[1+O(z)],
$$
and so this behaves like $\vpkoutp$ in (\ref{130}) provided that we take
$\sigma=+i|\sigma|$ (recall that $\sigma$ is pure imaginary). Then,
using (\ref{had}), the
Bogolubov coefficients can be read off directly from (\ref{grry}). Using
(\ref{normbog}), $W_{\lambda,\sigma}^*(z)=W_{-\lambda,\sigma}(-z)$, and the
relation
$$|\Gamma({1\over 2}+iy)|^2=\pi(\cosh\pi y)^{-1}$$
we have
\beq
|\beta_k|^2={\cosh\pi|\sigma-\lambda|\over
e^{2\pi|\sigma|}\cosh \pi|\sigma+\lambda|-\cosh\pi|\sigma-\lambda|}.
\label{alqi}
\eeq
For $|\sigma\pm\lambda|>>1$,
\beq
|\beta_k|^2\approx e^{-2\pi|\sigma+\lambda|},
\eeq
where
$$
|\sigma\pm\lambda|=H^{-2}\lc(\mm^2H^2+e^2E_0^2)^{1/2}\pm
eE_0\rc+O(H^2w^{-2}).
$$
As mentioned above, in keeping with standard notation we have taken $k>0$.
The result for $k<0$ is obtained by changing the sign of $e$ in the final
expression, since the differential equation only depends on the relative
sign of $k$ and $e$. Therefore
$$
|\beta_k|^2\approx
\exp\lp-{2\pi\over H^2}\lc(\mm^2H^2+e^2E_0^2)^{1/2}+eE_0\rc\rp,\quad (k>0)
$$
\beq
|\beta_k|^2\approx
\exp\lp-{2\pi\over H^2}\lc(\mm^2H^2+e^2E_0^2)^{1/2}-eE_0\rc\rp.\quad (k<0)
\label{nombre}
\eeq
In the instanton calculation we took the particle with charge $e$ to be the
one to the right of the inside region, and the particle with charge $-e$ to
be the one to the left. From (\ref{momentum}), the particle to the right has
$k<0$, so we have to use the second equation in order to compare with the
instanton results. From (\ref{se1}) and (\ref{nombre}), we have
$$
|\beta_k|^2\approx e^{-\bar S_E}.
$$
Therefore
\beq
{d\nn\over dx}=|\beta_k|^2{dk\over 2\pi}\approx e^{-\bar S_E}{dk\over
2\pi},
\label{jjj}
\eeq
in agreement with (\ref{compare2}).

Apart from the agreement between these momentum distributions and
between (\ref{compare4}) and (\ref{compared4}),
the time $t_k$ at which the definition of particle
starts being meaningful [see (\ref{timek})], is
the same as the time of nucleation $t_0$ in the instanton formalism
[see (\ref{momentum})], a suggestive coincidence.

\section{Summary and conclusions}

We have computed the nucleation rates for the process of membrane creation
by an antisymmetric tensor field in a spacetime of dimension $d=N+1$, for
$N=$1,2, and 3.

To this end, we have evaluated the contribution of the relevant instantons
to the semiclassical partition function. These instantons are $N$-spherical
worldsheets of radius $R_0$ [given by (\ref{r0})] embedded in a Euclidean
de Sitter background, which is itself a $d$-sphere of radius $H^{-1}$. The
flat space instantons are obtained by taking $H\to0$. We have discussed the
analytic continuation of the instantons, describing the motion of the
membranes after nucleation. The Lorentzian
solutions are spherical membranes which at late times expand like the scale
factor in the flat inflationary FRW model. The effect of
de Sitter transformations on a given solution
corresponds to space-time translations of the solution.

To evaluate the preexponential factor of the instanton contribution, it is
necessary to study small fluctuations of the instanton worldsheet. We have
reviewed the covariant theory of such perturbations, according to which the
normal displacement of the worldsheet is viewed as a scalar field $\phi$
living on the unperturbed worldsheet. We have given a kinematical
derivation of the equations of motion for $\phi$, based on the fact that
the zero modes, which correspond to infinitessimal translations of the
instanton, have to be solutions.

The evaluation of the prefactor is seen to be equivalent to the calculation
of the effective action for a free scalar field (the field $\phi$ mentioned
above) living in a curved background (the $N$-sphere). The functional
determinants that arise form Gaussian integration can be explicitly
calculated using $\zeta$-function regularization. This method automatically
removes all ultraviolet divergences. In the case $N=2$, an arbitrary
renormalization scale $\mu$ appears in the final result. Rescalings of
$\mu$ are seen to be equivalent to a finite renormalization of the membrane
tension  and of a new term which has to be added to the action. This new
term is just the Einstein-Hilbert action on the world-sheet, which for
$N=2$ is a topological invariant and hence does not contribute to the
equations of motion.

In flat space, the nucleation rates are obtained using the standard formula
which relates them to the imaginary part of the free energy [see
\ref{imf})]. We have recovered known results for the production of kinks in
$1+1$ dimensions, charged pairs in $3+1$ dimensions and `bubble' formation
in 2+1 and 3+1 dimensions. Our results apply only to the case when the
membranes are infinitely thin, having no internal structure. The effect of
finite thickness of the membrane can be important in realistic field
theories \cite{af79,afma82,AAM82,FSM84}. We plan to return to this question
in a forthcoming paper, where the thermal properties of the nucleated
objects are studied \cite{GVV93}.

In de Sitter space, there is no standard procedure for the calculation of
nucleation rates. We have introduced a heuristic prescription to obtain the
distribution of nucleated objects during inflation, following ideas related
to the treatment of this problem in Ref. \cite{baal91}. This distribution
is obtained from the one instanton contribution to the partition function.
The parameters corresponding to the zero mode rotations,
which form a subgroup of $O(d+1)$, have to be
analytically continued along with the instanton so that
they correspond to a subgroup of the de Sitter group $O(d,1)$.
With these manipulations, the one instanton contribution to the partition
function lends itself to interpretation as a distribution of membranes in a
space of parameters. The parameters can be chosen to be the place and
time of nucleation. In flat space this prescription reduces to the standard
formula (\ref{rate}) for the calculation of nucleation rates.
{}From the parameter distribution one can easily find the size
distribution of membranes in the inflationary universe, which turns out to be
stationary and nearly scale invariant [see (\ref{scaleinv})].

For the case of pair creation, we have given the distribution in terms of
the conserved coordinate momentum $k$.
This distribution turns out to be independent of
$k$ (as expected on the grounds of scale invariance) with an upper
cut-off $k^2_{phys}\approx \mm^2+(eE_0^2/H^2)$, where $e$ is the charge of
the particle, $E_0$ is the electric field, and $k_{phys}=e^{-2Ht}k$ is the
physical momentum. This cut-off corresponds to
the momentum of the particle at the time of nucleation $t_0$ (the physical
momentum is subsequently redshifted).

For the case of pairs, the calculations can be repeated using a
completely different approach. In 1+1 dimensional de Sitter space, the
Klein-Gordon equation for a charged scalar field coupled to a constant
electric field can be solved in terms of Whittaker functions. We find the
Bogoliubov transformations between an `in' state which is analogous to the
Bunch-Davies vacuum and an `out' state which corresponds to the definition
of particles at late times. This definition of particles at late times is
natural and unambiguous when the effective mass $\mm^2+(e^2E^2/H^2)$ is much
larger than the expansion rate (this is essentially why in everyday
experiments one does not care whether we live in flat space or in a de
Sitter space of very small $H$). In this limit, the results can be compared
with those obtained using the instanton formalism. Both methods agree, not
only in the exponential dependence but also in the prefactors. This is true
even when the size of the instantons is comparable to the horizon size.

To summarize, our results seem to indicate that the spontaneous nucleation
of defects during inflation and the decay of false or true vacuum through
nucleation of true or false vacuum bubbles is well described by the
instanton formalism, at least in 1+1 dimensions. Also, that our
prescription for finding the equilibrium distribution of nucleated objects
from the semiclassical partition function is correct. Although we have not
attempted to give a rigorous justification to this prescription, we hope
that with the examination of further examples a clearer picture will
emerge. After this paper was completed, we became aware of Refs.
\cite{lamo82}, dealing with the semiclassical approximation to the
path integral. The methods developed in these references may
be useful to provide a rigorous foundation to our prescription.
Work along these lines is currently under way.

Finally, some comments on negative modes. Coleman has shown \cite{co88}
that in flat space and at zero temperature an instanton describing the
decay of a metastable state has one and only one negative mode. As noted in
\cite{baal91}, the instantons describing the spontaneous nucleation of
defects during inflation can have more than one negative mode. Indeed,
we have seen that the scalar fields $\phi^{(i)}$ describing the transverse
displacements of the worldsheet are tachyonic, and that as a result
each field has a negative mode with $L=0$. Therefore, the number of
negative modes for such instantons is equal to the co-dimension of the
worldsheet. It was shown in \cite{baal91} that this is not in
direct contradiction
with Coleman's theorem, which does not apply in de Sitter space (or in flat
space at finite temperature). Still, the question remains of whether the
wrong number of negative modes renders the instanton `unphysical'.
{}From the analysis of the previous Section, we think that this is not the
case. Using the method of Bogoliubov transformations, we find that particles
are produced in de Sitter space for arbitrary co-dimensionality of the
world-line, and that the results are always in agreement with the instanton
results. Also, a nice feature of the prescription given in Section 7 is
that the distribution $d\nn$ is always real regardless of the co-dimension,
because the extra factors of $i$ coming from additional negative modes are
compensated by imaginary factors coming from the complexification of
additional boost zero modes.

\section*{Acknowledgements}

A am very grateful to Alex Vilenkin for stimulating conversations
and guidance during the preparation of this work. I would like to thank
Larry Ford, Diego Harari, Malcolm Perry, So-Jong Rey, Nami Svaitier
and Tanmay Vachaspati for numerous and interesting discussions,
and Ian Affleck for useful correspondence.
I am indebted to Julie Traugut for the layout of Fig. 1.
This work was partially supported by
the National Science Foundation, under contract PHY-9248784.

\section*{Appendix}
\setcounter{equation}{0}
\setcounter{section}{0}
\renewcommand{\theequation}{A\arabic{equation}}
\renewcommand{\thesubsection}{A\arabic{subsection}}

In this Appendix we evaluate the functional determinants for a free
scalar field $\phi$ on the $N$-sphere,
for $N=1,2$ and 3. As explained in section
\ref{perturbations} this scalar field represents small fluctuations of the
worldsheet of the instanton, and the determinant gives the pre-exponential
factor in the semiclasical partition function. The basic idea is to define
a generalized zeta function (\ref{zeta}) and then express the determinant
in terms of $\zeta(0)$ and $\zeta'(0)\equiv d\zeta/d z|_z=0$ [see Eq.
(\ref{detz})]. For the evaluation of $\zeta(z)$ we follow the
method of Ref.\cite{al83}, where the case $N=4$ was studied (for fields of
arbitrary spin).

\subsection{Determinant on the circle}

This is the case $N=1$. From (\ref{lambdal}) and (\ref{zdet}) the
eigenvalues are
$$
\Lambda_L=L^2+M^2R_0^2\equiv L^2+x^2, \quad (L=0,...,\infty).
$$
For $L\neq0$ the degeneracy is $2$ and for $L=0$ it is 1, so
$$
\zeta(z)=x^{-2z}+\sum_{L=1}^{\infty}2L^{-2z}\left(1+{x^2\over
L^2}\right)^{-z}.
$$
Expanding the binomial term in powers of $x/L$
\beq
\left(1+{x^2\over L^2}\right)^{-z}=\sum_{k=0}^{\infty}c_k\left({-x^2\over L^2}
\right)^k,
\label{binomial}
\eeq
we have
\beq
\zeta(z)=x^{-2z}+\sum_{k=0}^{\infty}2c_kx^{2k}(-1)^k\zeta_R(2z+2k),
\label{toexp}
\eeq
were $\zeta_R(z)\equiv\sum_{L=1}^{\infty}L^{-z}$ is the usual Riemann's
zeta function. Of the coefficients $c_k$ all we need to know is that
\beq
c_0=1,
\label{ck}
\eeq
$$
c_k={z\over k}+O(z^2).\quad (k\geq 1)
$$
Then it is clear that
$$\zeta(0)=1+2\zeta_R(0)=0,$$
since $\zeta_R(0)=-1/2$.

To evaluate (\ref{detz}) we also need $\zeta'(0)$. Expanding (\ref{toexp})
in powers of $z$ we have
$$
\zeta(z)=-2z\ln(2\pi x)+\sum_{k=1}^{\infty}2{z\over k}
x^{2k}\zeta_R(2k)(-1)^k + O(z^2),
$$
where we have used $\zeta_R(0)=-1/2,\zeta'_R(0)=-(1/2)\ln(2\pi)$. As a
result,
\beq
\zeta'(0)=-\ln(2\pi x)^2+\sum_{k=1}^{\infty}(-1)^k{2\over
k}x^{2k}\zeta_R(2k).\label{cuerno}
\eeq
Now we need a technique to sum the series.

For convenience one introduces the notation
$$
\zeta_R(z,\alpha)=\sum_{k=\alpha}^{\infty}L^{-z},
$$
so that $\zeta_R(z,1)=\zr (z)$. Using \cite{al83}
$$
{d^n\Psi(\alpha)\over d\alpha^n}=(-1)^{n+1}n!\zr(n+1,\alpha),
$$
one easily arrives at the expressions
\beq
\sum_{n=0}^{\infty}\zr(2n+1,\alpha)z^{2n}=-{1\over
2}[\Psi(\alpha+z)+\Psi(\alpha-z)],
\label{expression1}
\eeq
and
\beq
\sum_{n=1}^{\infty}\zr(2n,\alpha)z^{2n}={z\over
2}[\Psi(\alpha+z)-\Psi(\alpha-z)],
\label{expression2}
\eeq
where $\Psi(z)=d\ln\Gamma(z)/dz$ is the digamma function.

Now for the evaluation of (\ref{cuerno}). Differentiating with respect to
$x$ and using (\ref{expression2}) we have
$$
{d\over dx}\zz'(0)=-{2\over x}+2i[\Psi(1+ix)-\Psi(1-ix)].
$$
Upon integration we obtain the result
\beq
\zz'(0)=-2\ln(2\sinh\pi x).
\label{zpn1}
\eeq
In the last step, the constant of integration is chosen so that
(\ref{zpn1}) agrees with (\ref{cuerno}) when $x\to 0$. Restoring
$x=MR_0$ we obtain the desired result (\ref{z1sh}).

Note that, since $\zz(0)=0$ the determinant does not depend on the
arbitrary parameter $\mu$.
This was to be expected,
since after all for $N=1$
the path integral is just equivalent to the path integral
for a non-relativistic quantum mechanical oscillator at finite temperature.
That is, (\ref{int}) is equivalent to (\ref{epi}) with $V(x)=M^2x^2/2$ and
$\beta=2\pi R_0$. In (\ref{epi}) the right hand side is shorthand for
\beq
\lim_{n\to\infty}\int\prod_{i=1}^n{dx_i\over (2\pi\epsilon)^{1/2}}
\exp\left({-\epsilon\sum_i\left[{1\over
2}{(x_{i+1}-x_i)^2\over\epsilon^2}+V(x_i)\right]}\right),
\label{heavy}
\eeq
where $\epsilon=\beta/n$ and $x_{n+1}=x_1$. Here, there is no arbitrarity
in the definition of the measure of integration, and so we do
not expect any arbitrary scales to show up in the final result. The
integral (\ref{heavy}) can be done using the methods of Ref. \cite{geya60}
applied to the case of periodic boundary conditions, and one recovers the
result (\ref{z1sh}).

\subsection{Determinant on the 2-sphere}

In this case the eigenvalues are given by
$$
\Lambda_L=L(L+1)+M^2R_0^2\equiv (L+{1\over 2}+u)(L+{1\over 2}-u),
$$
where $u^2\equiv(1/4)-M^2R_0^2$. The degeneracies are given by $g_L=2L+1$.
It follows that
$$
\zz(z)=\sum_{n={1/2}}^{\infty}2n^{1-2z}\left[1-{u^2\over n^2}\right]^{-z},
$$
where $n$ runs over the positive half integers. Expanding the binomial term
in powers of $(u/n)$ one has
$$
\zz(z)=\sum_{k=0}^{\infty}2c_ku^{2k}\zr(2z+2k-1,{1\over 2}).
$$
Expanding in the neighborhood of $z=0$ we obtain
\beq
\zz(z)={1\over 12}+u^2+z[4\zr'(-1,{1\over 2})+Q]+O(z^2),
\label{corn}
\eeq
where
$$
Q\equiv\sum_{k=2}^{\infty}{2u^{2k}\over k}\zr(2k-1,{1\over
2})-2u^2\Psi\left({1\over 2}\right).
$$
Here we have used
$$
\zr(2z+1,\alpha)={1\over 2z}-\Psi(\alpha)+O(z),
$$
and the relation
$$
\zr(-1,\alpha)=-{1\over 2}\alpha^2+{1\over 2}\alpha-{1\over 12}.
$$
It is clear that $\zz(0)=(1/12)+u^2$.

To evaluate $\zz'(0)$ we need to find
$Q$. Using (\ref{expression1}) one easily arrives at
$$
{dQ\over du}=-2u[\Psi({1\over 2}+u)+\Psi({1\over 2}-u)].
$$
After a bit of algebra
$$
Q=-i\pi-3\ln\left({3\over 2}-u\right)+C+O(2u-3),
$$
where $C$ is a numerical constant of order unity. The last term, indicated
as $O(2u-3)$ vanishes when $u\to3/2$, i.e., when $M^2\to-2R_0^{-2}$, the
case we are interested in.

To summarize, we have
$$
\zz(0)={1\over 12}+u^2
$$
$$
\zz'(0)=-i\pi-3\ln({3\over 2}-u)+C'+O(2u-3),
$$
where we have absorbed the term $4\zr'(-1,1/2)$ in a new constant $C'$.
Using (\ref{ndetp}) one finds, in the limit $u\to
3/2$,
\beq
(det '[(\mu R_0)^{-2}\hat O])^{-1/2}=(\mu R_0)^{7/3}R_0^{-3}.\label{wak}
\eeq
Some numerical constants have been absorbed in a redefinition of the
renormalization scale $\mu$.

\subsection{Determinant on the 3-sphere}

In this case, from (\ref{lambdal}) and (\ref{zdet}), with $y^2\equiv
1-R_0^2M^2$,
$$
\Lambda_L=(L+1)^2-y^2.
$$
The degeneracy is given by $g_L=(L+1)^2$, so
\beq
\zz(z)=\sum_{n=1}^{\infty}n^{2-2z}\left[1-{y^2\over n^2}\right]^{-z}=
\sum_{k=0}^{\infty}c_ky^{2k}\zr(-2+2z+2k),
\label{yall}
\eeq
where $c_k$ are given by (\ref{ck}). Expanding around $z=0$ we have
$$
\zz(z)=\zr(-2)+z[2\zr'(-2)+Q]+O(z^2),
$$
where
\beq
Q\equiv\sum_{k=1}^{\infty}{y^{2k}\over k}\zr(2k-2).
\label{qe}
\eeq
It is clear that $\zz(0)=\zr(-2)=0$.

Also, $\zz'(0)=2\zr'(-2)+Q$. To evaluate $Q$ we first diferentiate
(\ref{qe}) and then use (\ref{expression2}) to find, after some algebra,
$$
{dQ\over dy}=-y^2{d\over dy}[\ln(\sin\pi y)].
$$
Integrating,
\beq
Q=-y^2\ln(\sin\pi y)+{2\over\pi^2}\int_0^{\pi y} x\ln(\sin x)dx,
\label{n}
\eeq
where the constant of integration is chosen so that $Q(y=0)=0$, in
order to agree with (\ref{qe}).

The integral in (\ref{n}) cannot be done analytically for
arbitrary $\pi y$. However, this is not a problem, since we want to
calculate the determinant only for $M^2\to-3R_0^{-2}$, which implies $y=2$.
Then the second term in (\ref{n}) is a definite integral which can be found
in the tables \cite{grry}, and we have
$$
Q(y)=-y^2\ln(\sin\pi y)-4\ln 2+3i\pi +O(y-2).
$$
Putting it all together we have
\beq
\zz(0)=0,
\label{wek}
\eeq
$$
\zz'(0)=2\zr'(-2)-4\ln 2+3i\pi -y^2\ln(sin\pi y)+O(y-2),
$$
where $y\equiv(1-R_0^{-2}M^2)^{1/2}$.

\section*{Figure caption}
\begin{itemize}
\item {\bf Fig.1} Euclidean de Sitter space is a
$d$-sphere of radius $H^{-1}$.
The instantons for membrane creation and
spontaneous nucleation of defects can
be seen as spherical worldsheets of dimension $N$
and radius $R_0$ embedded in
the $d$-sphere. For the case of spontaneous nucleation, the instantons have
maximal radius $R_0=H^{-1}$. For the case of membrane creation, the
co-dimension of the worldsheet is one, $d=N+1$, and the instanton can be
obtained by intersecting the $d$-sphere with a hyperplane at a distance
$\omega_0$ from the origin. The worldsheet of the membrane divides the
$d$-sphere into two regions, which we conventionally denote as the inside
and the outside of the membrane. The value of the electric field in the
outside region is taken to be equal to the background electric field before
nucleation $E_0$. By Gauss' law, the electric field in the inside region is
$E_0-e$.

\end{itemize}

\end{document}